\documentclass[10pt,journal]{IEEEtran}

\usepackage{hyperref}

\usepackage{algorithm}
\usepackage{algpseudocode}

\usepackage{times}
\usepackage{amsfonts}
\usepackage{amsmath}
\usepackage{verbatim}
\usepackage{epsfig}
\usepackage{enumerate}
\usepackage{xcolor,colortbl}
\usepackage{amssymb}
\usepackage{latexsym,color,amscd,textcomp,url}
\usepackage{rotating}

\usepackage{algorithm}
\usepackage{algpseudocode}
\usepackage{epsfig}
\usepackage{subfigure}
\usepackage{calc}
\usepackage{amsmath}
\usepackage{multirow}
\usepackage{adjustbox,lipsum}

\newcounter{countdefinitions}
\newcounter{counttheorems}
\begin{document}

\title{
CANTO - Covert AutheNtication with Timing channels over Optimized traffic flows for CAN
}

\author{\IEEEauthorblockN{ Bogdan Groza,  Lucian Popa and Pal-Stefan Murvay} \\
\thanks{Bogdan Groza, Lucian Popa and Pal-Stefan Murvay are with the Faculty of Automatics and Computers, Politehnica University of Timisoara, Romania,
Email:  bogdan.groza@aut.upt.ro, lucian.popa.lp@gmail.com, pal-stefan.murvay@aut.upt.ro}
}

\maketitle

\begin{abstract}

Previous research works have endorsed the use of delays and clock skews for detecting intrusions or fingerprinting ECUs on the CAN bus. Similar techniques have been also proposed for establishing a time-covert cryptographic authentication channel, in this way  cleverly removing the need for cryptographic material inside the limited payload of CAN frames. The main shortcoming of such works is the limited security level that can be achieved under normal CAN-bus traffic. In this work we endeavour to test the limits of the achievable security level by relying on optimization algorithms for scheduling CAN frames. Under practical bus allocations that are based on real-world scenarios, we are able to extract around 4--5 bits of authentication data from each frame which leads to an efficient intrusion detection and authentication mechanism. By accumulating covert channel data over several consecutive frames, we can achieve higher security levels that are in line with current security demands. To prove the correctness of our approach, we present experiments on state-of-the-art automotive-grade controllers (Infineon Aurix) and bus measurements with the use of industry standard tools, i.e., CANoe.
\end{abstract}

\section{Introduction and motivation}

Following small demonstrations for potential attacks on in-vehicle buses, e.g., \cite{Hoppe07}, the feasibility of attacking real-world vehicles has been proved by recent works such as \cite{Koscher10}, \cite{Checkoway11}, \cite{Miller13}, etc. All the security problems of in-vehicle networks stem from the fact that the Controller Area Network (CAN) is a decades-old bus which has no intrinsic security.

\begin{figure}[tb!]
\centering
\includegraphics[width=8.5 cm]{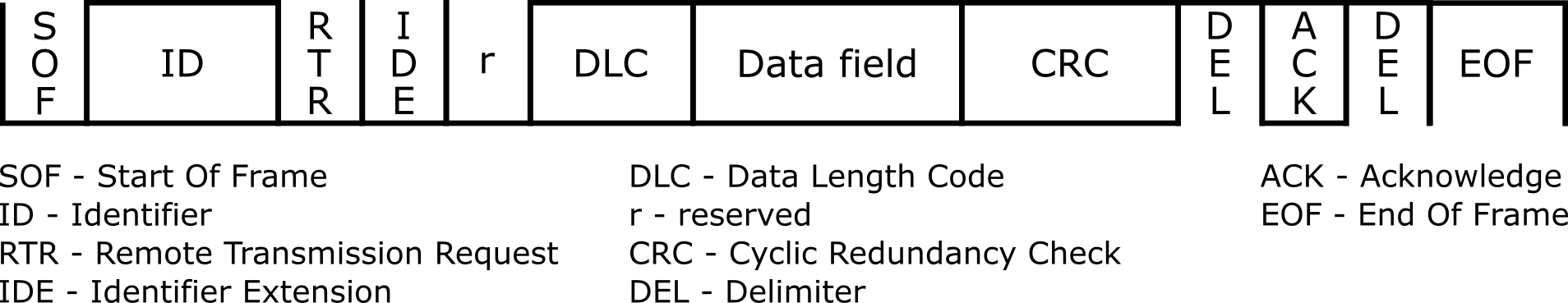}
\caption{Format of CAN data frames}
\label{fig:can_frm}
\end{figure}

CAN is the most widely employed  protocol for in-vehicle communications. It can transmit up to 8 bytes of data in a single frame (frame structure is depicted in Figure \ref{fig:can_frm}) at a maximum bit rate of 1Mbit/s. Being built with reliability in mind, CAN uses several mechanisms to assure this property. Arbitration is performed as an ID-based bus access priority to avoid collisions, while a 15 bit CRC aids in the identification of bit errors. For avoiding synchronization loss, CAN uses bit stuffing to ensure sufficient transitions in long sequences of bits of the same value. After each five consecutive identical bits, CAN introduces an additional bit of opposite value.

Following the recently reported attacks, the research community quickly answered with dozens of proposals for securing the CAN bus. As expected, most of these rely on the use of cryptographic Message Authentication Codes (MACs) (e.g.,  \cite{Hartkopp12}, \cite{Wang14}, \cite{Kurachi14},  \cite{Woo2016}, or more recently \cite{Bella19} and many others). But due to the limited size of the CAN frame, i.e., 64 bits, two options have been commonly discussed in the literature: truncating the MAC code or sending the MAC as a distinct packet. The last option creates additional problems since sending a new authentication frame for each regular frame doubles the bus-load and does not cope with practical demands. Nonetheless, it introduces authentication delays. The first procedure, MAC truncation, is supported by the more recent  security specifications introduced in the AUTOSAR \cite{AutosarSec} architecture which require 24 bit security for a CAN frame. However, reserving 24 bits out of the 64 bit CAN frame payload for security may not be convenient as this is represents 37\% of the payload. Additionally, the standard specifies 8 bits for a freshness parameter, leading to 32 bits reserved for security purposes and thus 50\% of the frame becomes unusable for regular data. In general, it seems that including cryptographic material in CAN frames remains somewhat problematic as the small packet size of CAN is hardly able to cope with the required level of security. A third option is to hide the authentication bits by using alternative physical layers such as CAN+ \cite{canplus} which is an extension of CAN. Such an approach was proposed in \cite{Herrewege11}. However, CAN+ transceivers do not exist inside vehicles and due to the migration to CAN-FD it seems unlikely for CAN+ to be adopted by the automotive industry.

\begin{figure*}[t!]
\centering
\includegraphics[width=16 cm]{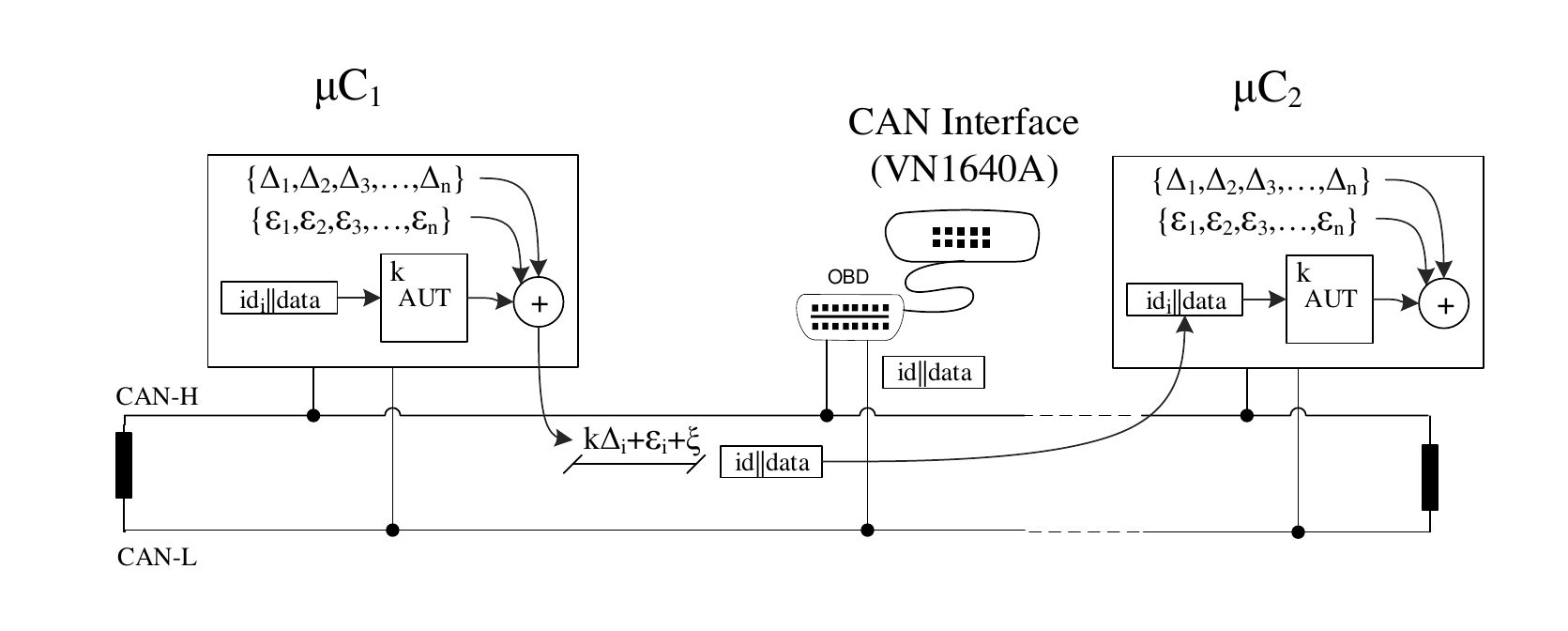}
\caption{Basic depiction of the addressed scenario: ECUs sending/receiving packets on the CAN bus, authentication data is encoded in delays}
\label{fig:scenario}
\end{figure*}

In a recent work, we exploited the fine-grained control of timer-counter circuits in a constructive manner by designing a time-covert cryptographic authentication channel \cite{Groza18}. This proposal has the merit of allowing authentication to be carried out outside the limited 64 bit payload of the CAN frame. However, the problem with the work in \cite{Groza18} is that performance degrades significantly when the covert channel is placed over existing (unoptimized) CAN-bus traffic. A similar approach for creating a covert authentication channel based on frame arrival time can be found in \cite{Ying19}, but the achieved security level is very limited at 1 covert bit for each CAN frame. Moreover,  the authentication in \cite{Ying19} is dedicated to the transmitter, not to the content of the frame itself.
On the other hand, covert timing channels have been well explored in computer networks, e.g., \cite{Berk05}, \cite{Liu09}, \cite{Cabuk04}, but we are unaware of the use of such channels for securing in-vehicle communication (except for the aforementioned recent papers \cite{Groza18} and \cite{Ying19}). We emphasize that, in contrast to such works, our intention is not in keeping the channel secret, but rather not interfering with the data-field and keeping the bus free. 

While there is not much related work on covert channels for CAN, there are several related works that are in close relation to our approach.
Optimal traffic allocation with respect to the security payload has been targeted by a small amount of works dedicated to CAN security such as \cite{Lin14}, \cite{Lin15},  \cite{Lin2015a} and \cite{Xie18}. These works do not target the creation of a covert timing authentication channel, but they focus on optimization problems for CAN traffic.
Nonetheless, many recent research works have been focusing on using frame arrival time, i.e., the delays that we use to create a covert channel, in order to detect intrusions, e.g., \cite{Moore17} and \cite{Song16}. By using Bloom filters \cite{Bloom70}, frame arrival time has been also combined with frame content to filter malicious activity in \cite{Groza18a}. More recently, frame periodicity  has been exploited to extract clock skews which is used to create a unique fingerprint for each ECU due to physical imperfections in oscillators in \cite{Cho16}. This sets room for physical fingerprinting of CAN nodes. The use of clock skews has been explored for fingerprinting computers for more than a decade by the work in \cite{Kohno05} and not surprisingly it was also applied to smart-phones \cite{Cristea13}.
Unfortunately, identification mechanisms based on clock-skews are rendered ineffective by the fine grained control of time-triggered interrupts on embedded devices which allows ECUs to potentially fake their clock-skews as demonstrated by \cite{Sagong18}. All these works are exploiting the precision of the clock circuitry in the controller, which also stays at the core of our proposal here. 

To save bits from the data-field, other works have suggested the use of the identifier field, i.e., \cite{Han15}, \cite{Humayed17},  \cite{Wu18} and \cite{Woo19}, but this requires special care as the identifier field is critical for arbitration and also used for filtering purposes.
An alternative to identify senders without compromising bits of the CAN frame is to use physical signal characteristics, e.g.,  \cite{Cho17}, \cite{Choi18}, \cite{Kneib18}, but these approaches may be vulnerable to small variations in bus impedance.

\begin{figure*}[t!]
\centering
\includegraphics[width=17cm]{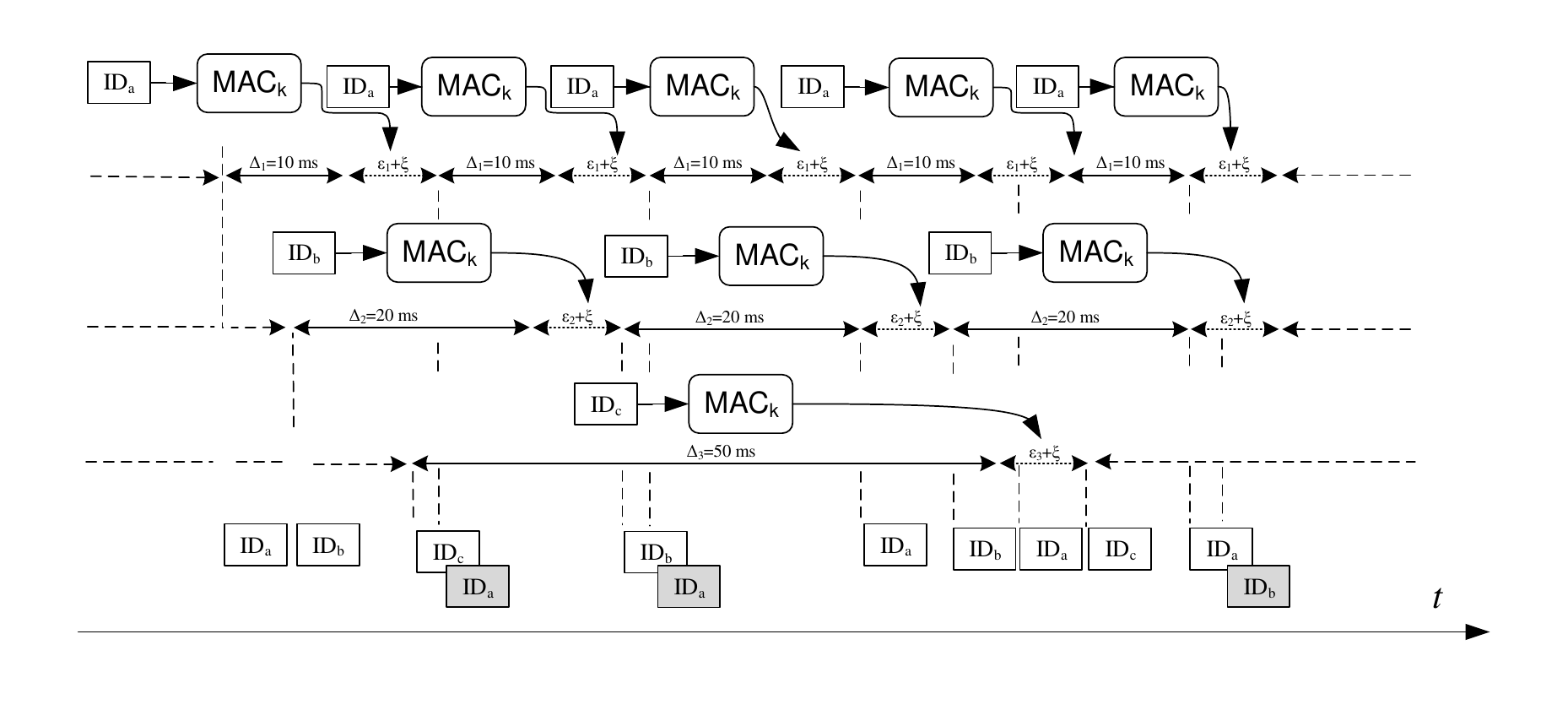}
\caption{Overview of the mechanism: frame arrival on the bus at delays $\Delta=10, 20, 50ms$ with adjustment $\epsilon_i, i=1..3$ and covert authentication delay $\xi$ resulting from a cryptographic MAC}
\label{fig:incanta}
\end{figure*}

\emph{Contribution in brief.} In this work we create a covert authentication channel, which leaves the bits in the CAN frame unchanged, and increase its data rate by relying on efficient frame scheduling. We do assume a bus-load of $\approx 40 \%$ which complies with recommendations for real-world implementations. We discuss several frame allocation mechanisms and find limits for the authentication payload that can be carried in a covert manner. The addressed scenario is briefly outlined in Figure \ref{fig:scenario}. To avoid overloading the figure, only two ECUs and one external device (potentially a CANcase) are depicted, but there are no restrictions regarding the number of ECUs or external devices in our scenario. The main advantages and disadvantages of a covert authentication channel on the CAN bus are the following:

\begin{itemize}
\item it does not consume bits from the data-field which is quite limited at 64 bits,
\item it covertly embeds authentication data in the frame that carries the data, without requiring an additional authentication frame,
\item it does not increase the bus-load since authentication data is hidden in delays.

\end{itemize}

The main concept behind how frame authentication works in our proposal, i.e., encoding authentication data in delays and adding optimizations for frame timings, is detailed in Figure \ref{fig:incanta}. CAN frames, depicted by the identifier field ID, arrive on the bus in a cyclic manner (to avoid overloading the figure, we omit the data-field, but this is used in the message authentication code along with the ID). While on-event frames may also exist on the CAN bus, the majority of the CAN traffic is cyclic in nature and we focus our work on authenticating such traffic. We depict identifiers for 3 distinct delays  $\Delta=10, 20, 50ms$. A drift $\xi$ is added to each delay which carries authentication data in a covert manner. In principle $\xi$ is the last byte of a cryptographic message authentication code (MAC). This MAC is computed over the content of the entire frame and will be distinct for every frame assuming proper use of freshness parameters, e.g., timestamps or counters. To avoid overloading the figure we omit such details in the graphical outline. Due to improper allocation of the CAN traffic, several packets may need to be transmitted at the same time (this is suggested by  packets highlighted in gray). Such schedule overlapping is not a problem from a transmission point of view, however, it impedes correct measurements of the arrival delays and thus the creation of a covert authentication channel. To avoid such situations, we use an additional delay $\epsilon_i, i=1..3$ in order to allocate traffic in an optimal manner and keep frame inter-distance at a maximum.

\begin{figure}[h!]
\centering
\begin{minipage}{4.25cm}
\includegraphics[width=4.25 cm]{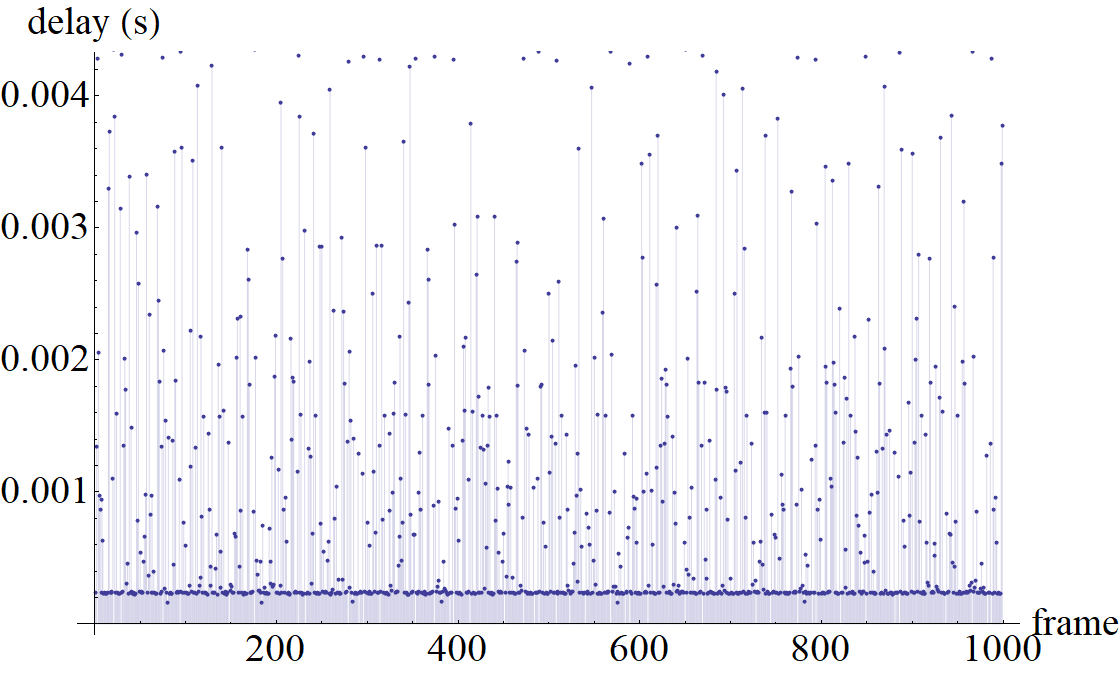}
\end{minipage}
\begin{minipage}{4.25cm}
\includegraphics[width=4.25 cm]{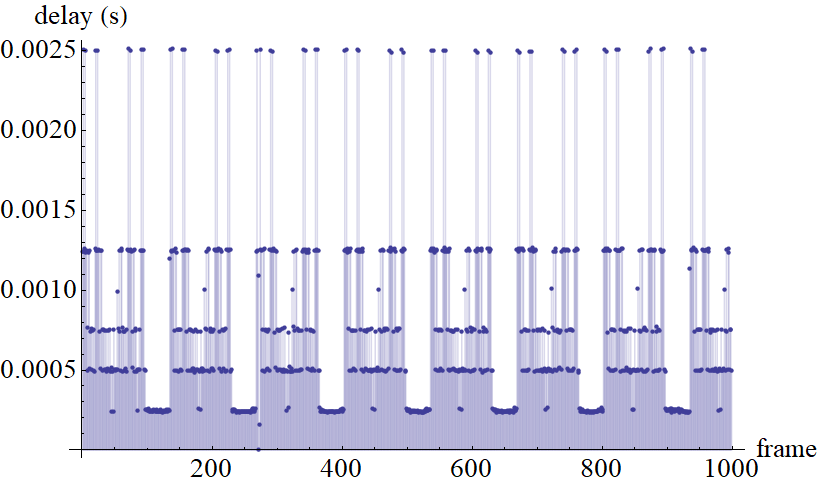}
\end{minipage}

\caption{Delays between frames: real-world car (left) vs. optimized traffic on our setup (right) at similar busload $30-40\%$}
\label{fig:inter_delay}

\centering
\centering
\includegraphics[width=7 cm]{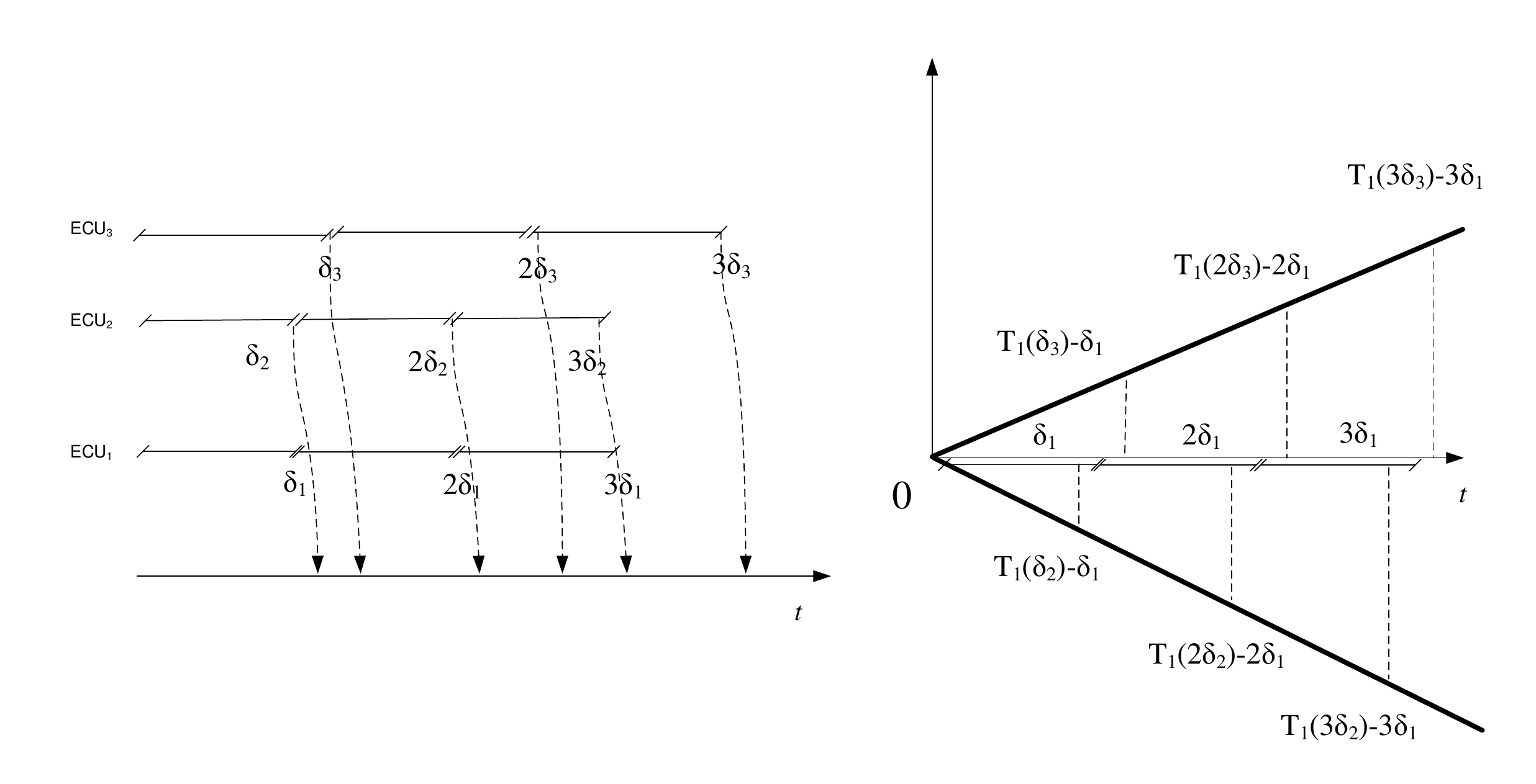}
\caption{Accumulation of clock skews for ECUs broadcasting at interval $\delta$ }
\label{fig:skew_new}

\centering
\centering
\includegraphics[width=7 cm]{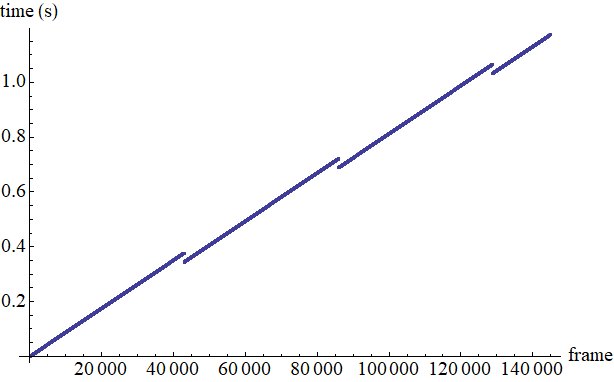}
\caption{Accumulation of clock skews for a device broadcasting at $100 \textit{ ms}$ for a longer period of $4$ hours }
\label{fig:skew_long}
\end{figure}

Figure \ref{fig:inter_delay} tries to clarify why unpredictable delays are problematic for a time-covert channel. The left side of the figure shows the inter-transmission times between frames as recorded in a real-world vehicle. While the entire traffic is cyclic, the inter-transmission time is noisy and deviations from the expected arrival time do greatly interfere with the creation of a time-covert channel. 
To improve performance we do rely on optimization algorithms. The right side of Figure \ref{fig:inter_delay} shows inter-transmission times after the traffic is optimized. The same bus-load and the same number of IDs is used, but the inter-transmission time now follows a clearer pattern. Creating a covert timing channel is possible without bus optimizations, but it is obvious that the optimized version will have a superior bitrate due to less noise on the covert channel.

\section{Background and experimental setup}

This section gives a brief overview on delays and clock skews on the CAN network. Nonetheless we discuss limitations in previous works on covert timing channels for the CAN bus. Then we describe the components of our setup.

\subsection{Clock skews and limitations in previous work}

In Figure \ref{fig:skew_new} we show how clock skews accumulate when three ECUs are broadcasting at fix time intervals $\delta$. While the delay $\delta$ is fixed, due to clock imprecision, the time measured at each ECU is in fact $\delta_1$, $\delta_2$ and $\delta_3$ respectively. If the first ECU measures the arrival time for frames received from the second and third ECUs, the delays accumulate. The result is a slope which represents the skew of the clock.

In Figure \ref{fig:skew_long} we show the clock skew from an experimental measurement over a longer period of 4 hours (since each frame is sent at $100 \mathit{ms}$ around $145.000$ frames are depicted). The slope of the line remains stable despite that the counter resets at roughly 45.000 frames resulting in a small drift. 

\begin{figure}[t!]
\centering

\begin{minipage}{4 cm}
\centering
\includegraphics[width=4 cm]{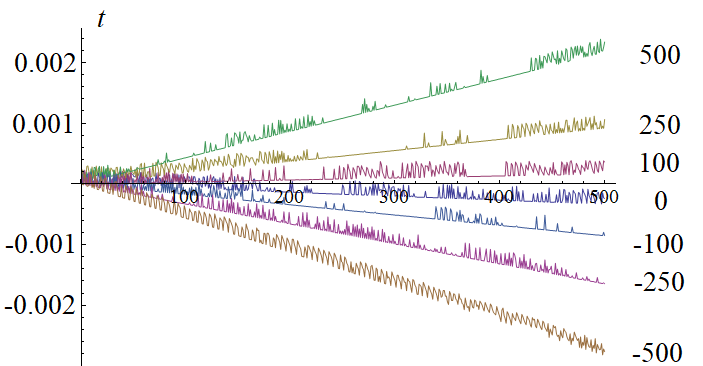}
\end{minipage}
\begin{minipage}{4 cm}
\centering
\includegraphics[width=4 cm]{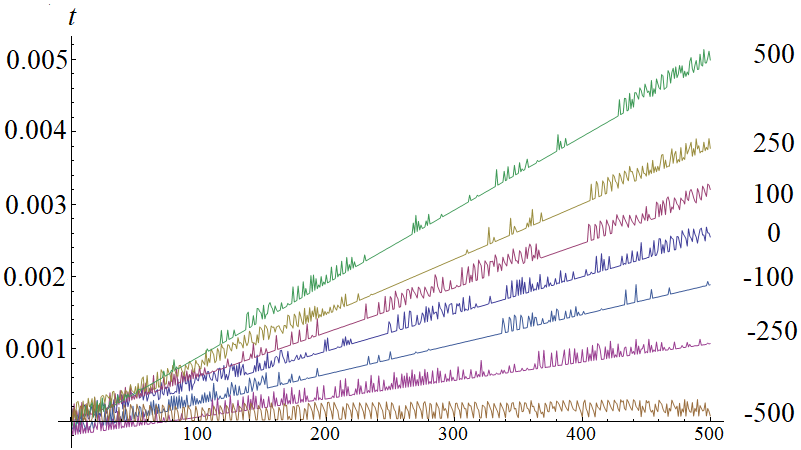}
\end{minipage}

\caption{Skews for a frame sent from an Infineon TC277 as recorded by an Infineon board (left) or from CANoe/VN CAN adapter (right)  in \cite{Groza18}}
\label{fig:skews277old}

\centering
\centering
\includegraphics[width=9 cm]{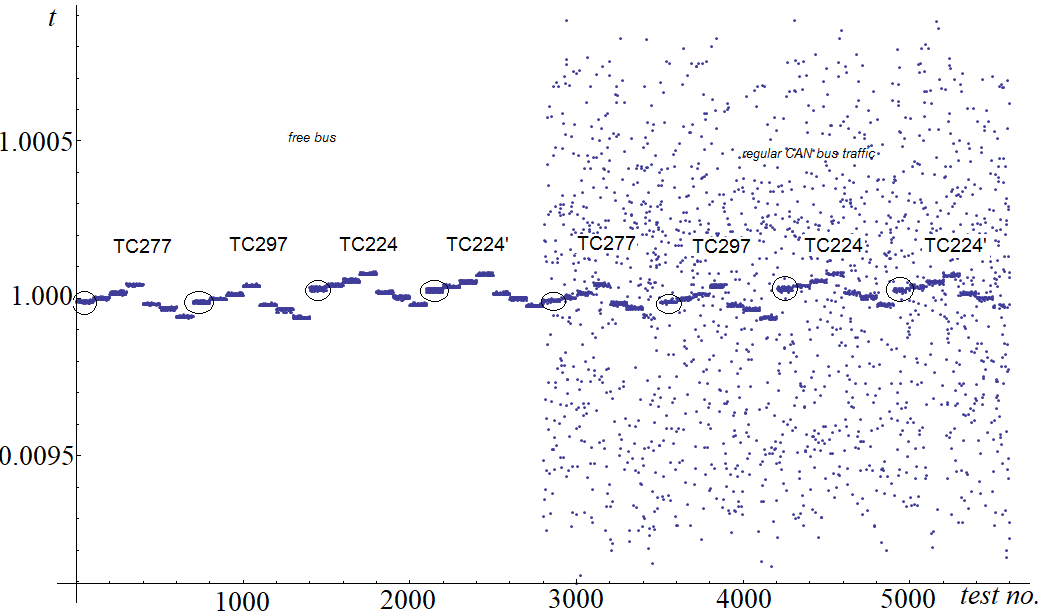}
\caption{Forced delays as recorded in \cite{Groza18} for a free bus (left) vs. a bus with regular network traffic (right)}
\label{fig:skew_all}
\end{figure}

Figure \ref{fig:skews277old} shows a graphical depiction for the delays measured on one Infineon board vs. CANoe in case of frames broadcast periodically by another Infineon board. The depiction is according to our previous work in \cite{Groza18}. Delays are forced at $\pm 100, \pm 250,\pm 500$ clock ticks ($1$ tick is $10ns$) and thus several slopes are visible in the picture.

The main limitation of our previous work on creating covert channels on the CAN bus was that existing traffic (poorly allocated) impedes the data-rate of the covert channel. Figure \ref{fig:skew_all} shows the variation of delays recorded on four Infineon TriCore boards  without (left) and with (right) existing network traffic according to \cite{Groza18} (delays are expressed as a fraction between the expected arrival time and recorded arrival time). In case of existing traffic, some of the frames arrive with significant delays making them indistinguishable for frames that are sent with random delays. These delays contribute to the false-positives of an intrusion detection mechanism. As we discuss and show in this work, traffic optimization is the only solution to this problem.

\subsection{Setup components}

\begin{figure}
\centering
\includegraphics[width=\columnwidth]{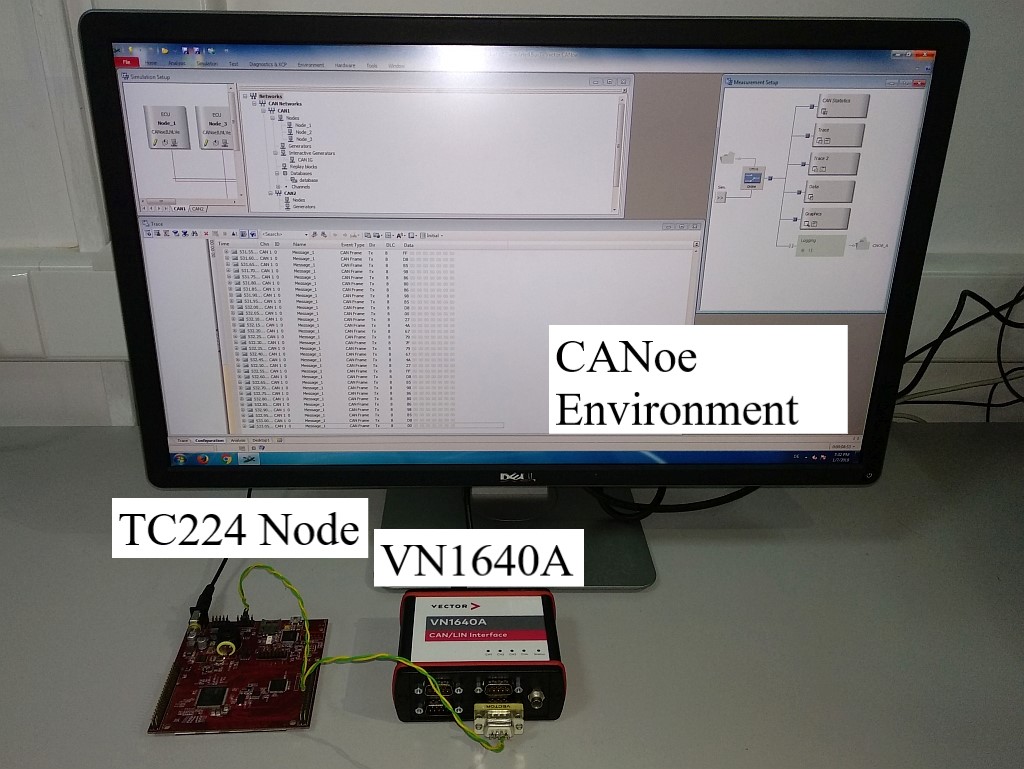}
\caption{Experimental setup used for generating and recording CAN traffic according to the proposed mechanisms}
\label{fig:exp_setup}
\end{figure}

We implement and evaluate optimizations on traffic allocation using an AURIX TC224 TFT Application Kit. The development board features a TC224 32-bit TriCore CPU that runs at frequencies up to 133 MHz and provides 1MB of FLASH memory and 96kB of RAM memory. The CAN frames transmitted by our TriCore-based implementation are recorded using CANoe, a software tool used for analyzing and testing of automotive networks. To achieve this, the CANoe running PC is interfaced with the development board through an VN CAN to PC adapter as depicted in Figure \ref{fig:exp_setup}. The recorded traces were analyzed offline using Mathematica.

Since, according to the described mechanism, CAN frames have to be transmitted in specific time slots, nodes need to implement a time keeping functionality. We implemented this on the TC224 using the Capture/Compare Unit 6 Timer (CCU6) module which was configured to trigger an interrupt every 1 ${\mu}$s as a base tick for our local clock. Additionally, we used the last 8 bits of a pre-computed MAC value as a delay, multiple of 1 ${\mu}$s based on the CCU6 Timer's ticks, between the message cycle time + ${\varepsilon}$ and the time when the message was actually sent on the bus, in the time-covert channel.

All of the message data bytes, configured message cycle times and the selected ${\varepsilon}$ values for each message were configured in the MultiCAN+ module. The MultiCAN+ module is also responsible for transmitting the frame data to the CAN transceiver with the specified baudrate of 500 kbps.

After performing the initial hardware setup, the initial MAC values are calculated for each message. During runtime, based on the counter value incremented in CCU6's timer, the cycle times for all frames and the ${\varepsilon}$ values, each frame is sent, but with a small delay as already described. The frame delivery will take place after ${\xi}$ ticks of the CCU6 Timer have expired. After each frame delivery, the message counter is incremented and a new MAC value is calculated based on the message data and the message counter.

\section{Optimizing traffic allocation}

This section addresses the optimization algorithms that we use. Traffic allocation is essential for achieving a satisfactory data-rate on the covert channel. We design and discuss four algorithms for traffic allocation and prove their effectiveness by both theoretical models/simulation and experimental data. We use two of these algorithms in the next section and implement the covert channel over optimized CAN-bus traffic. 

\subsection{Problem statement}

\newcommand\id{\mathsf{id}}
\newcommand\del{\Delta}
\newcommand\eps{\epsilon}
\newcommand\tst{T}
\newcommand\tmax{\delta^{\mathit{max}}_{\mathit{frame}}}

We consider a set of $n$ pairs $\{(\id_1, \del_1), (\id_2, \del_2),..., (\id_n, \del_n) \}$, each pair being formed by a CAN identifier and the delays (periodicity)  corresponding to the identifier. 
If on-event frames exist, a distinct mechanism should be used, this situation however is out of scope for our work. Further, let $\{(\id_i, \tst^i_1), (\id_i, \tst^i_2),..., (\id_1, \tst^i_l) \}$  be the set of identifier-timestamp pairs where timestamp $\tst^i_j, \forall j=1..l$ is the time at which $\id_i$ was received on the bus. Ideally, $\tst^i_{j+1} - \tst^i_{j} = \del_i, \forall i=1..n, j=1..l$, which means that frames having the same identifier ID are received at periodicity  $\del_i$. In practice however, there are many reasons that impede a perfect arrival time. Besides clock drifts, i.e., the clock of sender and receiver nodes is not identical, delays may occur due to frames with overlapped sending time. Since CAN arbitration is non-destructive, there is no problem if two nodes try to send a frame at the same time. But the frame with the higher ID loses arbitration and will be sent after the smaller ID which makes the arrival time drift from the expected $\del_i$.

\begin{figure}[t!]
\centering
\begin{minipage}{4.35cm}
\includegraphics[width=4.35 cm]{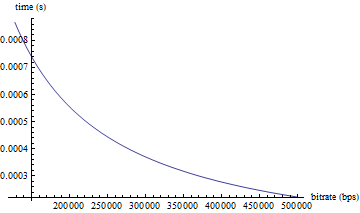}
\end{minipage}
\begin{minipage}{4.35cm}
\includegraphics[width=4.35 cm]{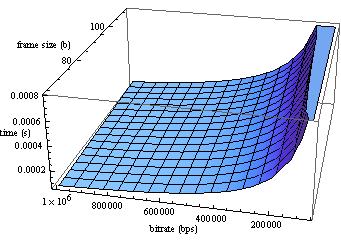}
\end{minipage}
\caption{Frame arrival time for a 64 bit data frame (left) and for a variable frame 0--64 bits (right) with data rates from 64kbps to 500kbps}
\label{fig:ftime}
\end{figure}

\emph{Frame arrival time.} The time required for a frame to be transmitted on the bus depends on the size of the frame and data rate of the bus. Datarate can be up to 1Mbps in standard CAN, though lower datarates of 125-500kbps are commonly employed. The size of the frame varies due to the number of stuffing bits, i.e., one bit of reverse polarity is added after 5 consecutive identical bits (for a frame with 64 bits of data plus the header, a maximum of 19 stuffing bits can be added). The left side of Figure \ref{fig:ftime}
shows the variation of frame arrival time in case of a 64 bits data frame which expands to 111 bits (without stuffing bits) and which may take as little as $100 \mu s$ on 1Mbps or up to $ 900 \mu s$ on a low-speed 125kbps bus (stuff bits not included). For a broader image, the right side of Figure \ref{fig:ftime} expands this calculation for variable size frames (0--64 bits) size and bus rates (64kbps--1Mbps).

\begin{figure}[t!]
\scriptsize

\centering
\begin{minipage}{4.25cm}
\includegraphics[width=4.25 cm]{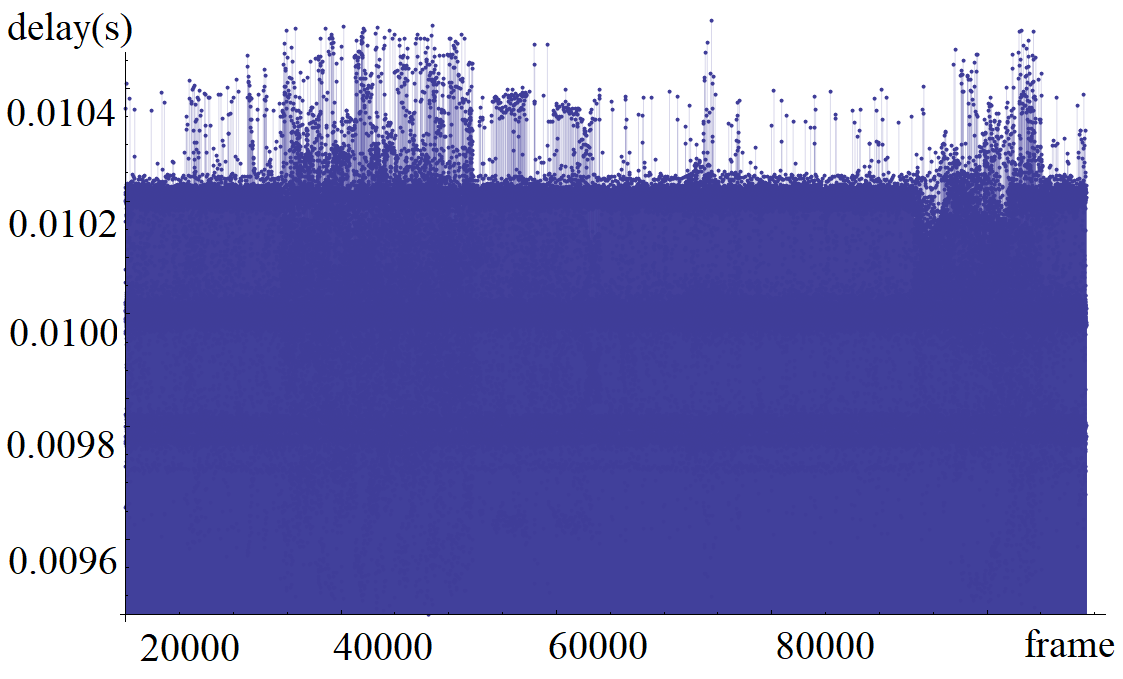}
\end{minipage}
\centering
\begin{minipage}{4.25cm}
\includegraphics[width=4.25 cm]{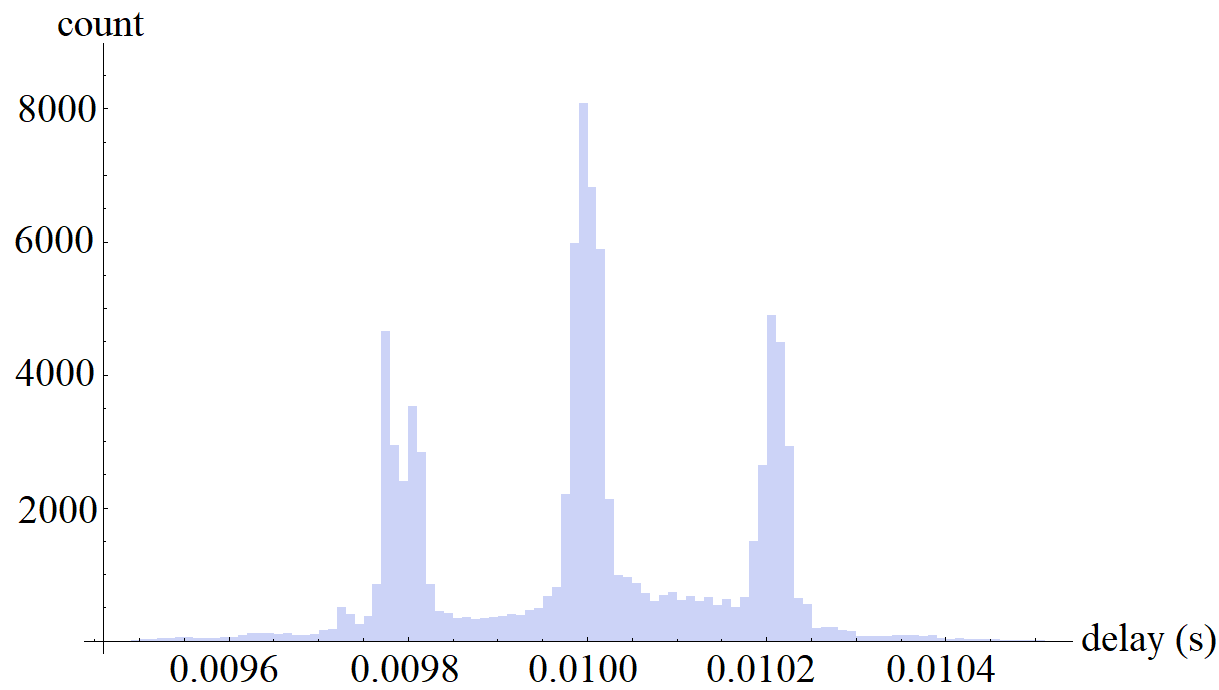}
\end{minipage}

(i)

\begin{minipage}{4.25cm}
\includegraphics[width=4.25 cm]{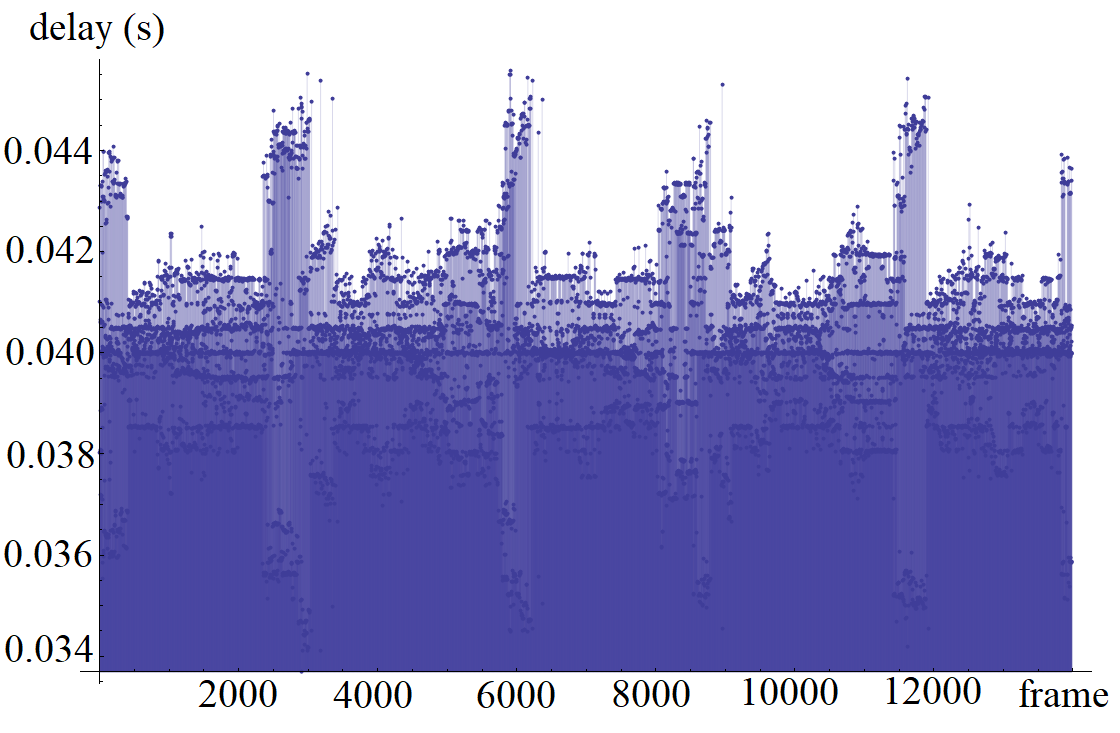}
\end{minipage}
\centering
\begin{minipage}{4.25cm}
\includegraphics[width=4.25 cm]{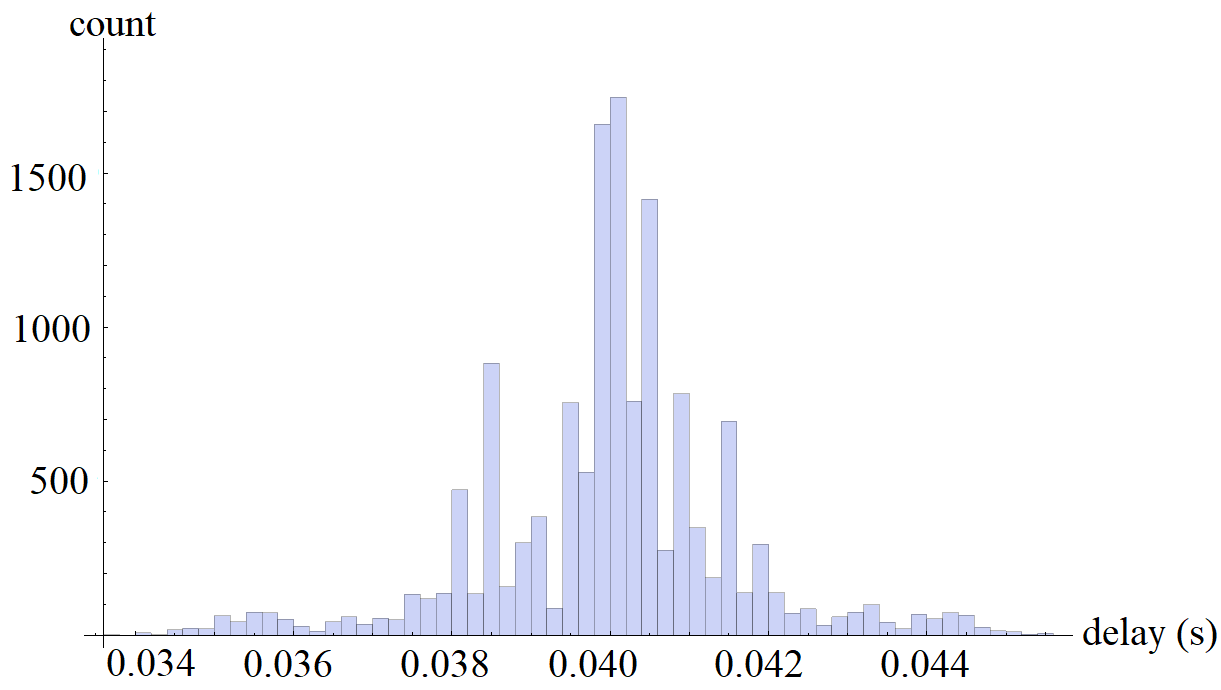}
\end{minipage}

(ii)

\begin{minipage}{4.25cm}
\includegraphics[width=4.25 cm]{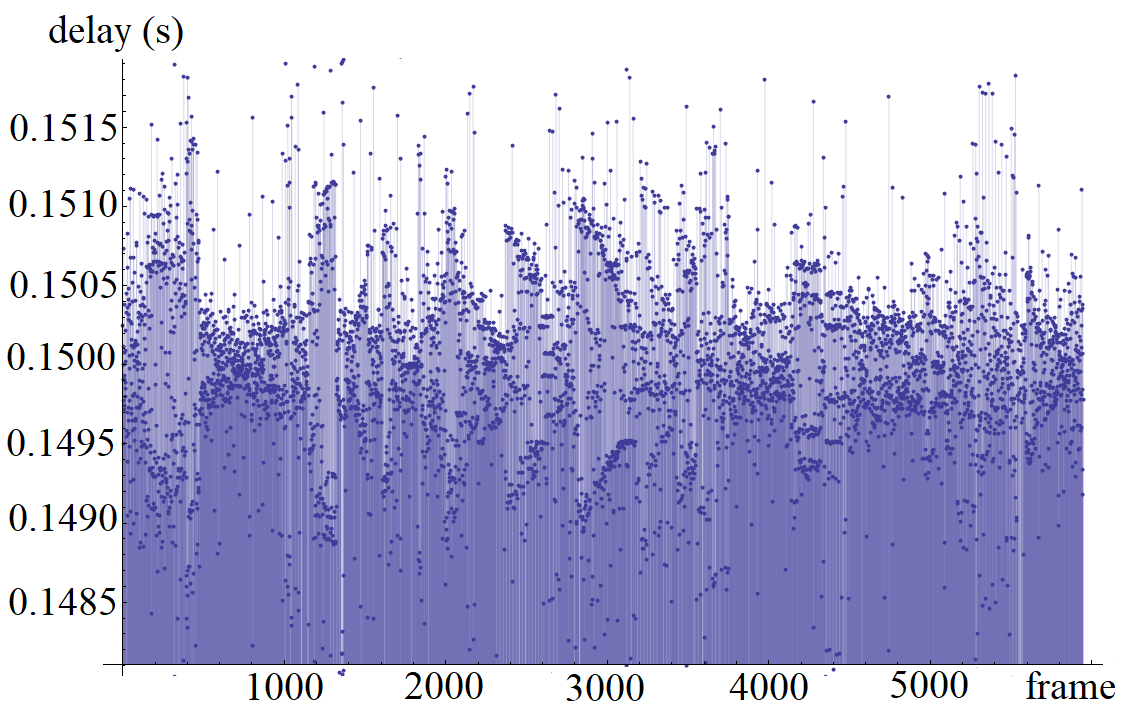}
\end{minipage}
\centering
\begin{minipage}{4.25cm}
\includegraphics[width=4.25 cm]{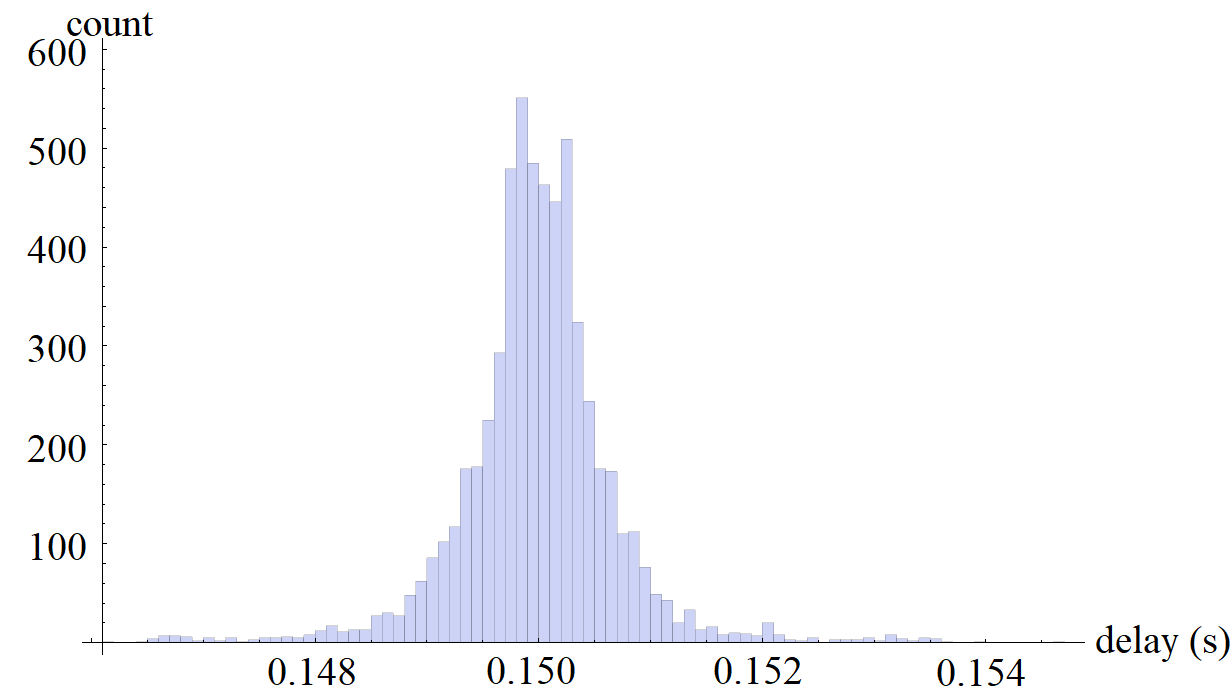}

\end{minipage}

(iii)

\begin{minipage}{4.25cm}
\includegraphics[width=4.25 cm]{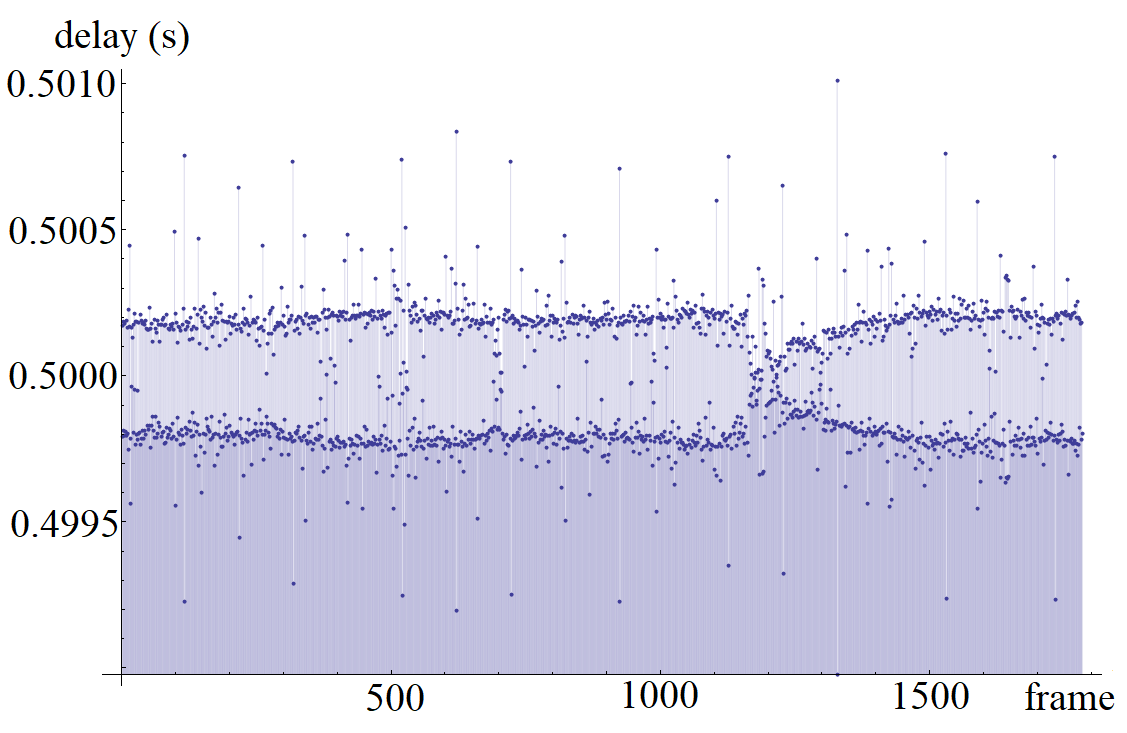}
\end{minipage}
\centering
\begin{minipage}{4.25cm}
\includegraphics[width=4.25 cm]{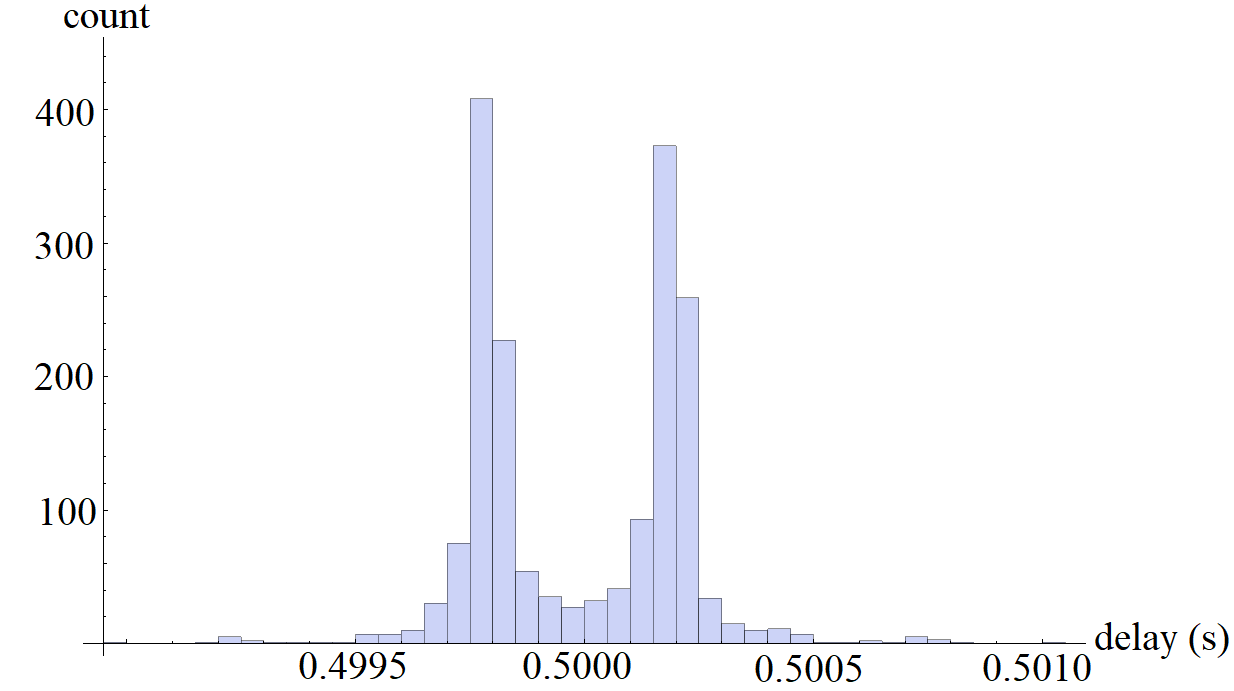}

\end{minipage}

(iv)

\caption{Frame arrival time and histogram distribution of frame arrival time, for frames arriving at 10ms, 40ms, 150ms and 500ms delays}
\label{fig:atime}
\end{figure}

\emph{Frame arrival time in real-world traces.} In Figure \ref{fig:atime} we depict the arrival time for frames scheduled at $10$, $40$, $150$ and $500 ms$. The left side of the figure shows the delay between frames carrying the same ID and the right side the histogram distribution of the same delay. Even for the higher priority frame arriving at $10ms$, deviations of $400 \mu s$ are common. For the $40 ms$ frame deviations of $2-4 ms$ are common and the situation is similar for the $150 ms$ frame. In case of the $500ms$ frame, deviations of $10ms$ become common as well. 

Such deviations from the expected arrival time exist and they clearly lower the bitrate of a covert timing channel.  The deviations from the expected arrival time are directly influenced by local clocks and the priority of the message ID, but these can be circumvented by clever allocation of frame timings as we discuss next.

\subsection{Optimizing frame scheduling}

In the previously defined framework, if each frame is sent at multiples of $\del_i, i=1..n$, the collisions on the bus between frame $i, j, \forall i,j =1..n$ will occur at multiples of $\mathit{lcm}(\del_i, \del_j)$ (here $\mathit{lcm}$ stands for the least common multiple of the two integers). This can be extended to any number of frames. In theory, all frames will collide on the bus at $\mathit{lcm}(\del_1, \del_2, ..., \del_n)$. Again, such collisions are non-destructive but they impede the time-covert channel. To avoid such collisions, we extend the frame scheduling set to $\{(\id_1, \del_1, \eps_1), (\id_2, \del_2, \eps_2),..., (\id_n, \del_n, \eps_n) \}$ where $\eps_i, i=1..n$ is a small drift added to the frame sending time. By our allocation, frames will be sent at intervals $k \del_i + \eps_i$ (rather than $k \del_i$). Our optimization problem consists in finding the values for $\eps_i, i=1..n$ such that for a given set of delays $\del_i, i=1..n$ no collisions will occur on the bus and moreover, the space between frames is maximized. 

We use the following theoretical model to compute optimal frame allocation. Let the following $n$ sets of traces corresponding to the $n$ IDs broadcast over the CAN network:

$$
\scriptsize
\begin{cases} 
T_1=\{(\id_1, \eps_1), (\id_2, \del_1 + \eps_1), (\id_3, 2 \del_1 + \eps_1), ..., (\id_1, (l-1) \del_1 + \eps_1) \} \\ 
T_2=\{(\id_2, \eps_2), (\id_2, \del_2 + \eps_2), (\id_3, 2 \del_2 + \eps_2), ..., (\id_2, (l-1) \del_2 + \eps_2) \} \\ 
... \\
T_n=\{(\id_n, \eps_n), (\id_n, \del_n +\eps_n), (\id_3, 2 \del_n + \eps_n), ..., (\id_n, (l-1) \del_n + \eps_n) \} \\ 
\end{cases}
$$

Having the previous equations, we say that the \emph{frame scheduling} is \emph{complete} if $|T_1 \cup T_2 \cup ... \cup T_n| = |T_1| + |T_2| + ... + |T_n|$ where $|T_i|, i=1..n$ denotes the cardinality of the set $T_i$. By this condition on the equality of the sets, we request that timings are distinct for all of the $n$ frames.

Let $T^{*} = \{t_1, t_2, ..., t_n \}$  be the set containing all time stamps for all messages, we assume this set to be sorted in ascending order, i.e., the natural way in which frames are expected to arrive on the bus. We say that the \emph{frame scheduling} is \emph{optimal} if the following value is minimal:

$$
q = \frac{1}{n} \sum_{i=2}^{n} \frac{1}{t_i - t_{i-1}}
$$

The quality factor $q$ in the previous equation is defined in order to assure a maximum inter-frame space (IFS). That is, the larger the IFS, the smaller the values $1/(t_i - t_{i-1})$ and thus their sum will be smaller. In what follows we discuss four variants of allocation algorithms that target the optimization of the quality factor $q$. 

\emph{A practical allocation example.} The subsequent optimization examples address the following frame periodicity vector which is based on existing CAN traffic from a real-world vehicle:

 $$
 \small
 \Delta = 
 \{
\underbrace{10, 10, ..., 10}_{\times 6}, 
\underbrace{20, 20, ..., 20}_{\times 8},  
\underbrace{50, 50, ..., 50}_{\times 12},  
\underbrace{100, 100, ..., 100}_{\times 14} 
\}
$$

That is, the vector contains 6 IDs that have a cycle time of $10 ms$, 8 IDs that have a cycle of $20 ms$, 12 IDs at $50ms$ and finally 14 IDs with a cycle time of $100 ms$

\newcommand\rdelays{\mathit{rdelays}}
\newcommand\blist{\mathit{blist}}
\newcommand\nblist{\mathit{blist'}}
\newcommand\leng{\mathit{size}}
\newcommand\appe{\mathit{append}}
\newcommand\dele{\mathit{delete}}

\begin{figure}
\begin{center}

\begin{minipage}[thb]{9cm}
 \vspace{0pt} 
\begin{algorithm}[H]
\small
\caption{Binary symmetric allocation}
\label{alg:bin_sym_alg}
\begin{algorithmic}[1]
\Procedure{Binary allocation}{}
 \State $\rdelays \leftarrow \{ \del_2, ..., \del_n \}$ 
 \State $\blist \leftarrow \{ (0, \del_1) , (\del_1, \infty) \}$ 
  \While{$\leng(\rdelays) > 0$}
  \State $\nblist \leftarrow \{\blist[1]\}$
 \For{$i=2, i \leq \leng(\blist) $}
 \If{$\leng(\rdelays)>0$}
 \State $\nblist \leftarrow \appe (\nblist,(\blist[i-1, 1] $ 
 \Statex ~~~~~~~~~~~~~~~~~~~~~~~~~~~~~~~~~~~~~~~~~~~~~~~~~~~~~~$ - \blist[i, 1])/2, \rdelays[1]))$ 
 \State $\nblist \leftarrow \appe (\nblist, \blist[i])$
 \State $\rdelays \leftarrow \dele (\rdelays, 1)$
 \Else
 \State $\nblist \leftarrow \appe (\nblist, \blist[i])$
 \EndIf
 \EndFor
 \State $\blist \leftarrow \nblist$
 \EndWhile
\EndProcedure
\end{algorithmic} 
\end{algorithm}
\end{minipage}

\end{center}
\caption{Algorithm for binary symmetric allocation}
\label{fig:bin_sym_alg}

\centering
\begin{minipage}{4.25cm}
\includegraphics[width=4.25 cm]{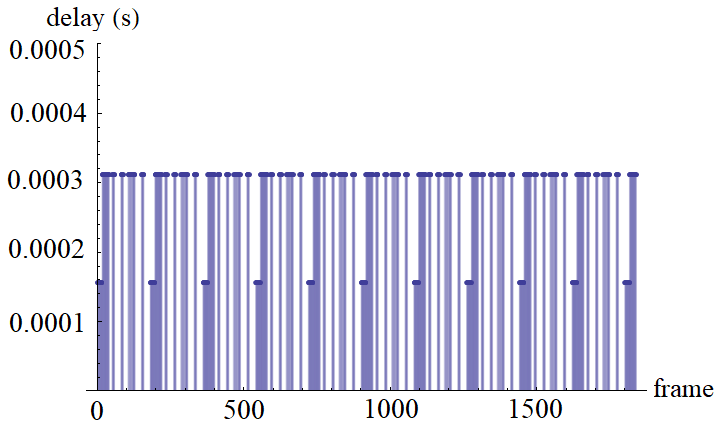}
\end{minipage}
\begin{minipage}{4.25cm}
\includegraphics[width=4.25 cm]{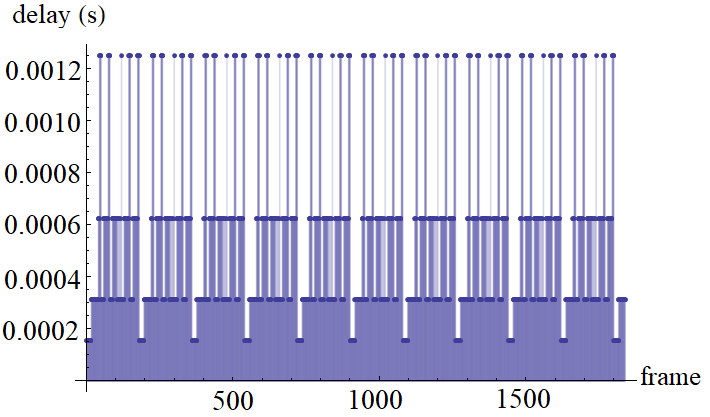}
\end{minipage}
\caption{Theoretical displacement of delays in case of binary symmetric allocation: detail for delays lower than $300 \mu s$ and overall view up to $1.2 ms$ for the first 2000 frames }
\label{fig:bin_alloc}

\centering
\begin{minipage}{4.25cm}
\includegraphics[width=4.25 cm]{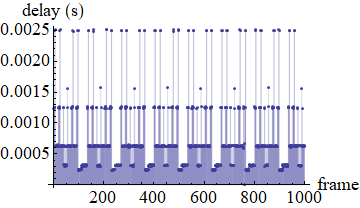}
\end{minipage}
\begin{minipage}{4.25cm}
\includegraphics[width=4.25 cm]{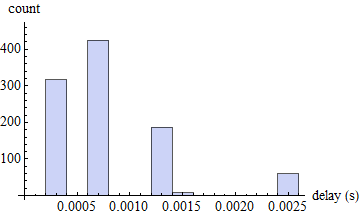}
\end{minipage}
\caption{Experimental measurements from CANoe of an Infineon node broadcasting after binary symmetric allocation: delays (left) and  histogram distribution of delays (right)}
\label{fig:bin_alloc_exp}
\end{figure}

\emph{Binary symmetric allocation}. This is the simplest of the allocation algorithms that we use, it is very easy to implement and gives good results (we improve however on the inter-frame space with the next algorithms). In the binary symmetric allocation, we start with a bin size equal to the window size $w$ which is the minimum of the delays $w = \min{(\del_1, \del_2, ..., \del_n})$ and allocate all the values $\epsilon_i, i=1..n$ to fit symmetrically in the interval $[0..w]$. For this we start with a bin of size $w$ then we create new values $\epsilon_i$-s by dividing each existing bin. That is, $\epsilon_1$ is first placed at $w/2$, then for $\epsilon_2$ and $\epsilon_3$ two new bins are created at $w/4$ and $3w/4$, etc. 
The algorithm is presented in Figure \ref{fig:bin_sym_alg}. We assume the delays appear in $\rdelays$ in ascending order ($\del_1$ is missing from the list since it is already allocated to a default value $\epsilon_1=0$). Variable $\blist$ is instantiated with the first delay $\del_1$ (for which we allocate a default value $\epsilon_0=0$) and a second delay $\infty$ which is merely a placeholder for delimiting the bin which has size $\del_1$. In step 4 we loop until an $\epsilon_i$ is generated for each value $\del_i$. For this purpose, we create a new list $\nblist$ in step 5 and loop in step 6 over all existing values in $\blist$ to create a new value $(\blist[i-1, 1] - \blist[i, 1])/2$ that is added between $\blist[i-1]$ and $\blist[i]$ in the newly created list $\nblist$. In step 15 $\blist$ is replaced with $\nblist$ at each iteration. At the end of the procedure $\blist$ will contain all pairs $\{ (\epsilon_1, \del_1), ..., (\epsilon_n,\del_n) \}$.
A graphical depiction of the theoretical frame timings is in Figure \ref{fig:bin_alloc}. In Figure \ref{fig:bin_alloc_exp} we present the experimental results as measured from CANoe in when an Infineon node broadcasts frames with the corresponding timings. The theoretical results and experimental measurements are reasonably close. Differences exists as several frames are broadcast later resulting in a $2.5 ms$ inter-frame delay. The reason for this is that the minimum inter-frame space is at $250 \mu s$ which is also around the time needed to place a frame on the bus at 500 kbps. Due to computational delays on the controller, if the time-slot is missed, the frame will be sent at some later point missing the expected allocation on the bus. We conclude that $250 \mu s$ for inter-frame space is somewhat too short. 

\newcommand\delays{\mathit{delays}}
\newcommand\sort{\mathit{sort}}
\newcommand\rem{\mathit{remove}}
\newcommand\opt{\mathit{opt}}
\newcommand\ind{\mathit{ind}}
\newcommand\epses{\mathit{reps}}

\begin{figure}
\begin{center}

\begin{minipage}[thb]{9cm}
 \vspace{0pt} 
\begin{algorithm}[H]
\small
\caption{Randomized allocation}
\label{alg:rand_alloc_alg}
\begin{algorithmic}[1]
\Procedure{Randomized Allocation}{}
 \State $e \leftarrow \min(\del_1, \del_2, ..., \del_n)/n$
 \State $\epses \leftarrow \{ 0, e, 2e, ..., (n-1)e \}$ 
 \State $\delays \leftarrow \{ \del_1 ,\del_2, ..., \del_n) \}$ 
 \State $i=1$
 \State $q \leftarrow \infty$
 \State $\epses'' \leftarrow \perp$
 \While{$i \leq \mathit{max_{iterations}} $}
 \State $\epses' = \mathit{randomize}(\epses)$
 \State $T^* \leftarrow \{\} $
 \For{$j=1, i\leq n, j=j+1$}
 \State $T_j \leftarrow \{ k\del_j + \epses'[j]; k=1..\lfloor T/\del_j \rfloor \} $
 \State $T^* \leftarrow \appe(T^*, T_j) $
 \EndFor
 \State $T^* \leftarrow \sort(T^*) $
 \State $q'  = \frac{1}{n}\sum_{i=2}^{|T^*|} \frac{1}{t_i - t_{i-1}} $ 
 \If{$q'<q$}
 \State  $\epses'' = \epses' $
 \State  $q = q'$
 \EndIf
\EndWhile
\EndProcedure
\end{algorithmic} 
\end{algorithm}
\end{minipage}

\end{center}
\caption{Algorithm for Randomized search}
\label{fig:rand_alloc_alg}

\centering
\begin{minipage}{4.25 cm}
\includegraphics[width=4.25 cm]{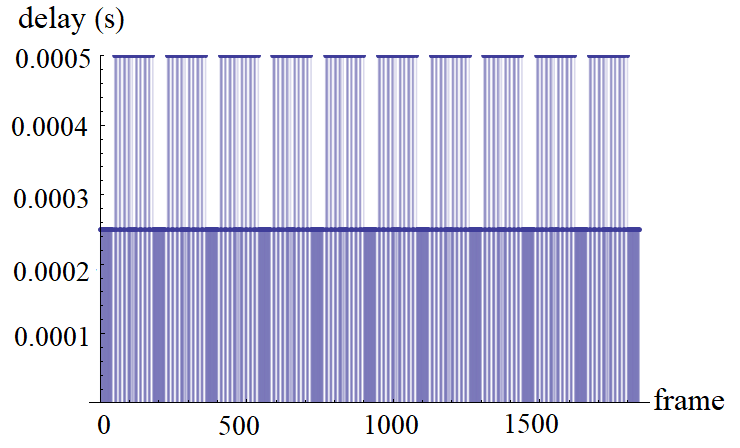}
\end{minipage}
\begin{minipage}{4.25 cm}
\includegraphics[width=4.25 cm]{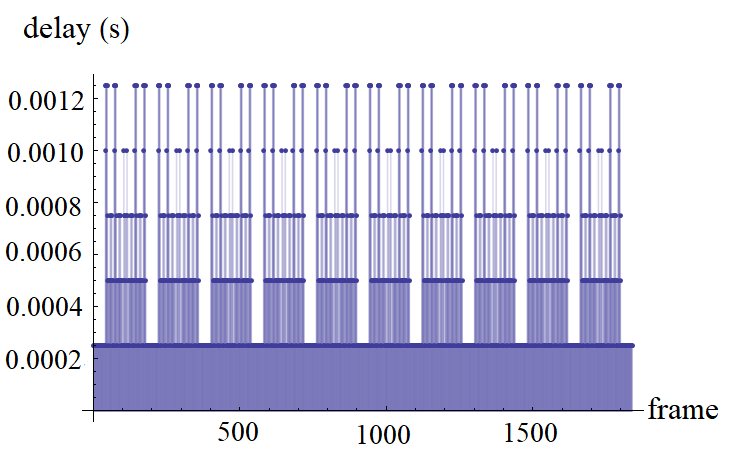}
\end{minipage}
\caption{Theoretical displacement of delays in case of Randomized allocation: detail for delays lower than $300 \mu s$ and overal view up to $1.2 ms$ for the first 2000 frames }
\label{fig:rand_alloc}

\centering
\begin{minipage}{4.25cm}
\includegraphics[width=4.25 cm]{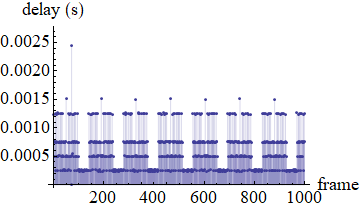}
\end{minipage}
\begin{minipage}{4.25cm}
\includegraphics[width=4.25 cm]{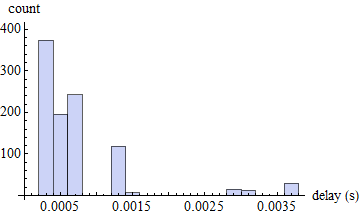}
\end{minipage}
\caption{Experimental measurements from CANoe of an Infineon node broadcasting after Randomized allocation: delays (left) and  histogram distribution of delays (right)}
\label{fig:rand_alloc_exp}
\end{figure}

\emph{Randomized search allocation.} The randomized search first creates a list of $\epsilon_i, i=1..n$  that are equally spaced. A distance set to $e$ where $e = \min{(\del_1, \del_2, ..., \del_n)/n}$ is a natural choice since at worst all frames will appear during the periodicity of the fastest frame from the bus. However, other values may be fixed for $e$. Then the $\epsilon_i, i=1..n$ values are allocated randomly to each delay $\del_i, i=1..n$. After $\ell$ iterations (each consisting in a randomized allocation) the best allocation is kept. The result improves with a higher number of iterations. 
The algorithm is presented in Figure \ref{fig:rand_alloc_alg}. First, the value of $e$ is computed in line 2, then values of the target epsilons are generated in line 3 and stored in $\epses$. The current optimum $q$ is set to $\infty$ in line 6 and the best allocation $\epses''$ is set to void in line 7. Then we loop in line 8 for $\mathit{max_{iterations}}$. During each loop, the new values $\epses'$ are set to a random permutation from $\epses$ in line 9. Lines 10--16 compute the quality factor $q$ for the new allocation. For this, the set of timestamps $T_j, j=1..n$ is generated for each delay, according to the corresponding $\epsilon_j$ of the current permutation. The timestamps are computed for a timeframe $T$ (in our practical tests we set this to 1-10 seconds). Then the set of timestamps $T^*$ is sorted and $q$ is computed accordingly. If the result is better than for the previous optimum, i.e., line 17, the new result is stored in $\epses''$, otherwise it is dropped.
The randomized search gave somewhat better results than the binary symmetric allocation, but again the $250 \mu s$ seems to be problematic for some frames (we improve on this with the next two algorithms). Figure \ref{fig:rand_alloc} holds the theoretical expectations and Figure \ref{fig:rand_alloc_exp} holds the experimental measurements, these are again reasonably close to each other.

\begin{figure}[thb!]
\centering
\begin{minipage}{4.25cm}
\includegraphics[width=4.25 cm]{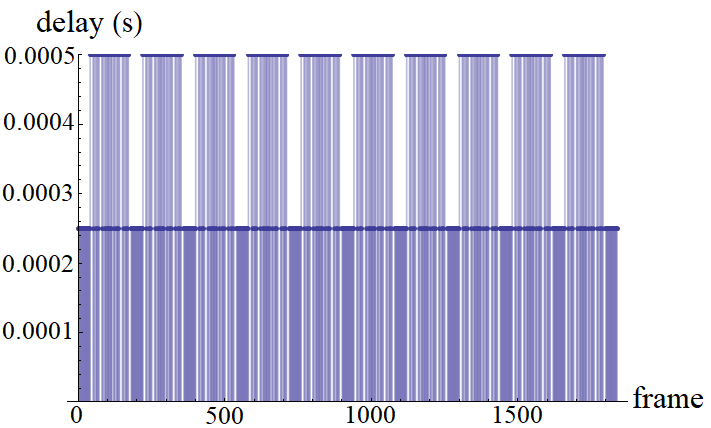}
\end{minipage}
\begin{minipage}{4.25 cm}
\includegraphics[width=4.25 cm]{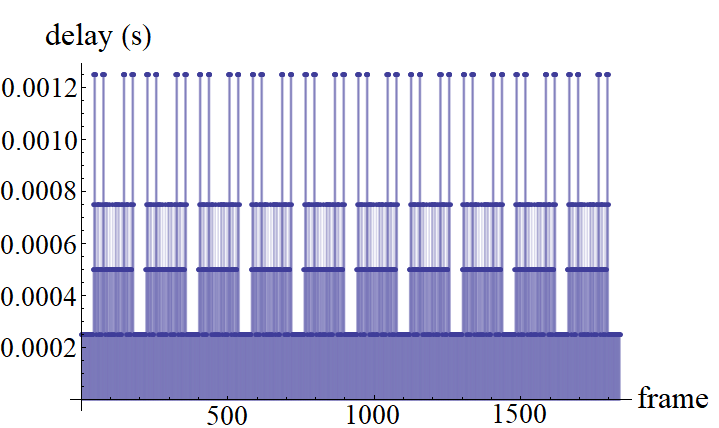}
\end{minipage}
\caption{Theoretical displacement of delays in case of Greedy allocation: detail for delays lower than $300 \mu s$ and overal view up to $1.2 ms$ for the first 2000 frames }
\label{fig:greedy_alloc}

\centering
\begin{minipage}{4.25cm}
\includegraphics[width=4.25 cm]{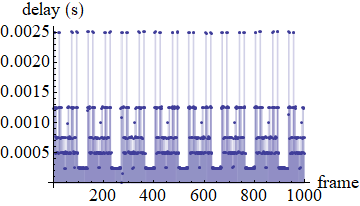}
\end{minipage}
\begin{minipage}{4.25cm}
\includegraphics[width=4.25 cm]{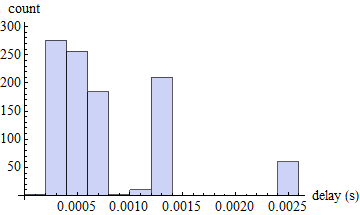}
\end{minipage}
\caption{Experimental measurements from CANoe of an Infineon node broadcasting after Greedy allocation: delays (left) and  histogram distribution of delays (right)}
\label{fig:greedy_alloc_exp}

\centering
\begin{minipage}{4.25cm}
\includegraphics[width=4.25 cm]{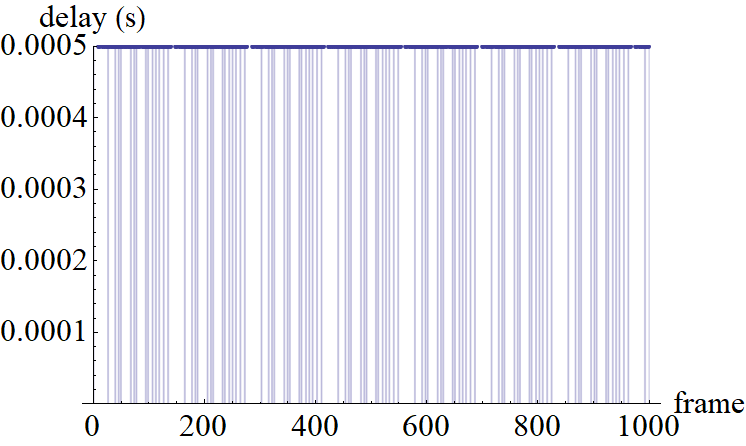}
\end{minipage}
\begin{minipage}{4.25 cm}
\includegraphics[width=4.25 cm]{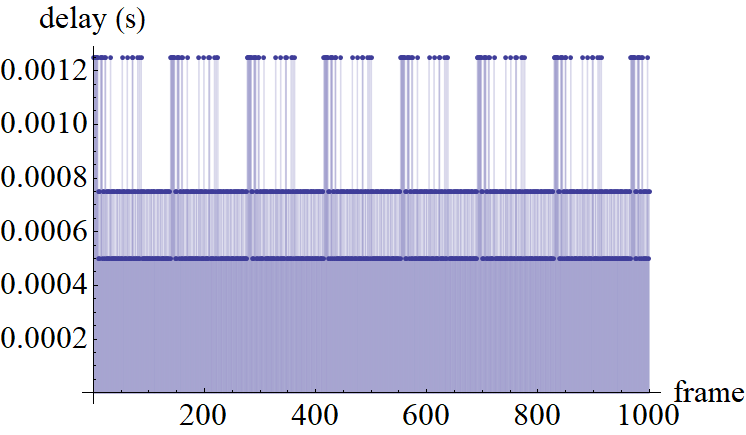}
\end{minipage}
\caption{Theoretical displacement of delays in case of Multi-Layer Greedy allocation: detail for delays lower than $300 \mu s$ and overall view up to $1.2 ms$ for the first 2000 frames }
\label{fig:mgreedy_alloc}

\centering
\begin{minipage}{4.25cm}
\includegraphics[width=4.25 cm]{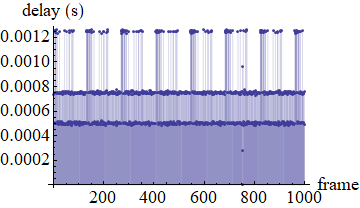}
\end{minipage}
\begin{minipage}{4.25cm}
\includegraphics[width=4.25 cm]{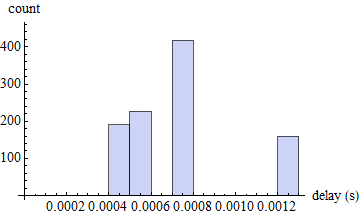}
\end{minipage}
\caption{Experimental measurements from CANoe of an Infineon node broadcasting after Multi-Layer Greedy allocation: delays (left) and  histogram distribution of delays (right)}
\label{fig:mgreedy_alloc_exp}
\end{figure}

\emph{Greedy allocation}. In the Greedy allocation, for $n$ delays, we first create bins  that are equally spaced at distance $e = \min(\del_1, \del_2, ..., \del_n)/n$ (this is identical to the case of the previous algorithm). Then we allocate the delays in ascending order in such way that $q$ is minimized. 
The algorithm is summarized in Figure~\ref{fig:greedy_alg}. The algorithm starts by building the set of values $\epsilon_i, i=1..n$ in a similar way to the randomized allocation described previously. Then it loops for each delay in line 7 and for each of the remaining values $\epsilon$ loops again in line 10 in order to find the optimum value for the current delay. Each of the selected values is tested against the optimization criteria in a similar manner to the randomized allocation. The index of the optimal value is stored in $\ind$ and this is removed from the $\epsilon$ values stored in $\epses$ in line 21 such that only the remaining values could be allocated for the next delay $\del_i$.
Figure \ref{fig:greedy_alloc} and \ref{fig:greedy_alloc_exp} hold the theoretical expectations and experimental measurements. The results are closer than for the previous two algorithms which suggests Greedy to be a better optimization.

\emph{Greedy Multi-Layer.} To finally circumvent the $250 \mu s$ inter-frame space issues we modify the Greedy allocation to a Multi-Layer Greedy allocation in which frames at delay $\del_i, i=1..n$ are allowed to be placed at any multiple of $e$ that is smaller than $\del_i$. This allows for a better expansion of the frames since frames at a larger $\del_i$ can benefit from a larger $\epsilon_i$. We skip formalism for this algorithm to avoid overloading the paper. The theoretical and experimental results are in Figures \ref{fig:mgreedy_alloc} and \ref{fig:mgreedy_alloc_exp}. This time the results are almost identical and the inter-frame space is expanded to up to $500 \mu s$. 

\begin{figure}
\begin{center}

\begin{minipage}[thb]{9cm}
 \vspace{0pt} 
\begin{algorithm}[H]
\small
\caption{Greedy allocation}
\label{fig:greedy_alg}
\begin{algorithmic}[1]
\Procedure{Greedy Allocation}{}
 \State $e \leftarrow \min(\del_1, \del_2, ..., \del_n)/n$
 \State $\epses \leftarrow \{ 0, e, 2e, ..., (n-1)e \}$ 
 \State $\epses' \leftarrow \{  \}$ 
 \State $\delays \leftarrow \{ \del_1 ,\del_2, ..., \del_n) \}$ 
 \State $T^* \leftarrow \{\}$
 \For{$i=1, i \leq \leng(\delays), i=i+1 $}
 \State $\opt = \infty$
 \State $\ind = 1$
 \For{$j=1, j \leq \leng(\epses), j=j+1 $}
\State $T^*_{\mathit{aux}} \leftarrow T^*$
\State $T_j \leftarrow \{ k\del_j + \epses'[j]; k=1..\lfloor T/\del_j \rfloor \} $
 \State $T^*_{\mathit{aux}} \leftarrow \appe(T^*_{\mathit{aux}}, T_j) $
 \State $T^*_{\mathit{aux}} \leftarrow \sort(T^*_{\mathit{aux}}) $
 \State $\opt'  = \frac{1}{n}\sum_{i=2}^{T^*_{\mathit{aux}}} \frac{1}{t_i - t_{i-1}} $ 
 \If{$\opt'<\opt$}
 \State  $\ind = j$
 \State  $\opt = \opt'$
 \EndIf
 \EndFor
 \State $\epses \leftarrow \rem(\epses, \ind)$
 \State $\epses' \leftarrow \appe(\epses', \ind)$
 \State $T_i \leftarrow \{ k\del_i + \epses'[i]; k=1..\lfloor T/\del_i \rfloor \} $
 \State $T^* \leftarrow \appe(T^*, T_i)$
 \EndFor
\EndProcedure
\end{algorithmic} 
\end{algorithm}
\end{minipage}

\end{center}
\caption{Algorithm for Greedy allocation}
\label{fig:greedy_alg}
\end{figure}

\begin{figure}[thb!]
\centering
\begin{minipage}{4.25cm}
\includegraphics[width=4.25 cm]{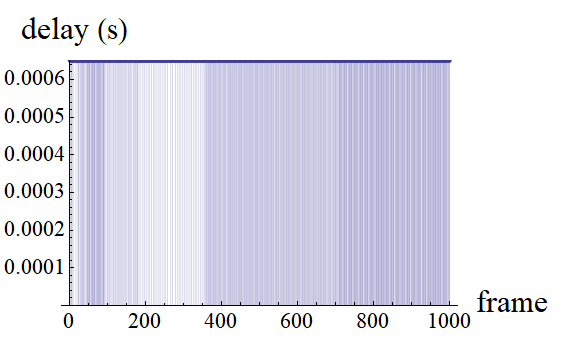}
\end{minipage}
\begin{minipage}{4.25 cm}
\includegraphics[width=4.25 cm]{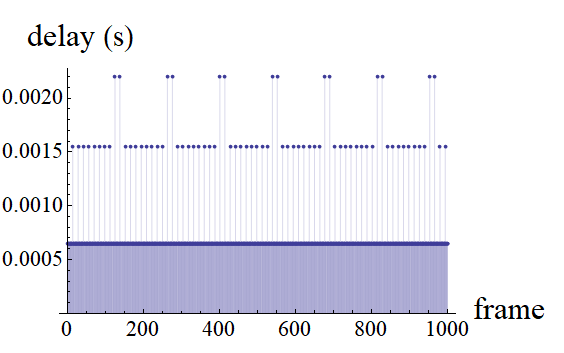}
\end{minipage}
\caption{Theoretical displacement of delays in case of GCD at $0.6ms$ allocation: detail for delays lower than $300 \mu s$ and overal view up to $1.2 ms$ for the first 2000 frames }
\label{fig:gcd_alloc}

\centering
\begin{minipage}{4.25cm}
\includegraphics[width=4.25 cm]{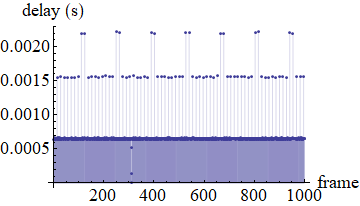}
\end{minipage}
\begin{minipage}{4.25cm}
\includegraphics[width=4.25 cm]{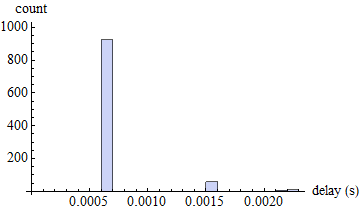}
\end{minipage}
\caption{Experimental measurements from CANoe of an Infineon node broadcasting after GCD allocation at $0.6ms$ inter-frame space: delays (left) and  histogram distribution of delays (right)}
\label{fig:gcd_alloc_exp}
\end{figure}

\emph{GCD based allocation}. We also try a Greatest Common Divisor (GCD) based allocation in which the frames are spaced by a delay, e.g., fixed at $500 \mu s + \delta_i$ where $\delta_i, i=1..n$, which is a multiple of $\gcd(\del_1, \del_2, ..., \del_n)$ subject to the condition that $500 \mu s + \delta_i$ is smaller than $\del_i$. 
The algorithm is depicted in Figure \ref{fig:gcd_alg}. It starts from building a matrix $M$ of $\max(\delays)/G$ rows and $\min(\delays)/\epsilon$ columns, where $G$ is the greatest common divisor of the delays and $\epsilon$ is the minimum allowable IFS. The matrix is initially filled by 1s. Then in step 7 the algorithm loops for all the delays searching for each of them an empty row in the matrix $M$. The search is done in the loop from step 11 starting from row $j=1$ (initialized in step 9) and continuously increases $j$ in step 35 until a row without 0s is found. To test that the current row has non-zero values, the loop in step 14 goes until the end of the current line $aux$. If the resulting $k$ is smaller than the length of the line $|aux|$, then starting from step 17, the line is filled with zeros at $l=d/G$ steps (where $d$ is the current delay). The result for the current delay is placed in $\epses$ in line 31.
The allocation provides results similar to the case of the Multi-Layer Greedy allocation, the runtime of the algorithm is however much faster. A potential drawback is that the algorithm doesn't search for minimizing $q$, it rather places frames at a minimum IFS fixed to $\epsilon$. The theoretical and experimental results are in Figures \ref{fig:gcd_alloc} and \ref{fig:gcd_alloc_exp} and we attempt to increase the inter-frame space to $600 \mu s$ which we succeed. 

In Table \ref{tab:comp} we summarize a comparison between the four optimization algorithms. 
The main drawback with the first three algorithms is that they leave a minimum inter-frame space of only $150 \mu s$ - $250 \mu s$ which is problematic since it may be smaller than the transmission window for a single frame (this may cause delays that pile-up and compromise the data-rate of the covert channel). The multi-layer Greedy and circular GCD show better performance and the minimum IFS is kept at $500 \mu s$ which sufficient for the transmission of single frames.

\newcommand\aux{\mathit{aux}}
\newcommand\mat{\mathit{M}}

\begin{figure}
\begin{center}

\begin{minipage}[thb]{8cm}
 \vspace{0pt} 
\begin{algorithm}[H]
\small
\footnotesize
\caption{GCD-based allocation}
\label{fig:gcd_alg}
\begin{algorithmic}[1]
\Procedure{GCD-based Allocation}{}
  \State $\delays \leftarrow \{ \del_1 ,\del_2, ..., \del_n) \}$ 
  \State $G \leftarrow \gcd (\delays)$ 
  \State $\epsilon \leftarrow 0.5 $
  \State $\epses \leftarrow \perp $ 
  \State $M \leftarrow 1^{[1..\max(\delays)/G][1..\min(\delays)/\epsilon]}$
 \For{$i=1, i \leq \leng(\delays), i=i+1 $}
 \State $d \leftarrow \delays[i]$
 \State $j = 1$
 \State $s = \mathit{false}$
 \While{$s=\mathit{false} \& j \leq |\mat|$}
 \State $k = 1$ 
 \State $\aux = \mat[j]$
 \While{$k \leq |\aux| \&  \aux[k] = 0$}
 \State $k=k+1$
 \EndWhile
 \If{$k \leq |\aux|$}
  \State $l = d/G$
  \State $a = k$
  \State $s = \mathit{true}$
  \While{$a \leq |\aux|$}
   \If{$\aux[a] \neq 0$} 
   \State $\aux[a] = 0$
   \State $a = a + l$
   \Else
   \State $s = \mathit{false}$
   \EndIf
  \EndWhile
  \If{$s = \mathit{true}$}
  \State $\mat[j] = \aux$
  \State $\epses = \appe(\epses, (j-1) \epsilon)$
  \EndIf
 \EndIf
 \EndWhile
 \State $j=j+1$
 \EndFor
\EndProcedure
\end{algorithmic} 
\end{algorithm}
\end{minipage}

\end{center}
\caption{Algorithm for GCD-based allocation}
\label{fig:gcd_alg}
\end{figure}

\begin{table}[b!]
\begin{center}
\caption{Comparison of allocation algorithms}
\label{tab:comp}
\begin{tabular}{ c c c c c } 
\hline
\hline
    & Completeness & q-Factor & Min IFS & Max IFS \\
\hline
\hline
 Binary Sym. Search& \checkmark & 2.37 & 0.15 & 2.5  \\
 Randomized Search & \checkmark  & 2.51 & 0.25 & 1.25   \\
 Greedy Search & \checkmark  & 2.39 & 0.25 & 1.25   \\
 Greedy ML Search & \checkmark  & 1.50 & 0.5 & 1.1   \\
 Circular GCD & \checkmark  &  1.86 & 0.5 & 1.0   \\
  \hline
\end{tabular}
\end{center}
\end{table}

\section{Protocol and results}

In this section we give an outline of the protocol, then we discuss practical results on the covert timing channel that carries authentication tags. 

\newcommand\tol{\rho}

\subsection{Main protocol}

\newcommand\send{\mathsf{SendCyclic}}
\newcommand\receive{\mathsf{RecCyclic}}

\newcommand\ttag{\mathit{tag}}
\newcommand\mac{\mathit{MAC}}
\newcommand\key{\mathsf{k}}
\newcommand\mes{\mathrm{m}}
\newcommand\intr{\mathsf{Intrusion}}
\newcommand\erraut{\mathsf{AutFailed}}
\newcommand\delay{\Delta}
\newcommand\timer{\mathsf{T}}
\newcommand\ctime{\mathsf{t}}
\newcommand\wait{\mathit{wait}}
\newcommand\err{\epsilon}
\newcommand\slev{\ell}

In the current work we are focused on bus optimization for achieving a maximum capacity for the covert channel. The protocol for sending and receiving frames is not distinct from our previous proposal \cite{Groza18} named INCANTA since we use the same kind of covert channel based on the drift of the frame from the expected arrival time. A minor difference is that to each delay $\del_i, i=1..n$ we add the corresponding value $\eps_i, i=1..n$ resulting from the optimization algorithm. For consistency, we keep the same name and description for our protocol INCANTA (INtrusion detection in Controller Area Networks  with Time-covert cryptographic Authentication). INCANTA consists in the following set of actions that are to be followed by each node:

\begin{enumerate}

\item 
$\send(\id_i, \mes)$  is the procedure triggered at some fixed delays $k\del_i +\eps_i$ for a frame with identifier field $\id_i$ at which the responsible sender ECU computes the tag $\ttag = \mac_\key(k, \id_i, \mes)$ where $k$ is a counter that is incremented for each new message to be send. The sender then sets $\timer=\lfloor \ttag \rfloor_{\slev}$ and performs a wait operation $\wait(\timer)$ then broadcasts message $(\id, \mes)$,

\item 
$\receive(\id_i, \mes)$ at which the $k^{\mathit{th}}$ instance of a message with identifier $\id_i$ is received. Let time $\ctime_k$ be the time at which the message is received, the receiver computes $\ttag = \mac_\key(k, \id, \mes)$ and $\timer_k=\lfloor \ttag \rfloor_{\slev}$ then checks if $|\ctime_k-\ctime_{k-1}|-(delta+\timer_{k}-\timer_{k-1})| \leq \tol$ and if this fails it drops the frame and reports an intrusion otherwise it considers the frame as genuine.

\end{enumerate}

We consider that a shared secret key $\key$ exists on each ECU from the CAN  bus. We do not discuss how this key is shared since this is addressed by several other works. Moreover, we consider that the frame-scheduling set $\{(\id_1, \del_1, \eps_1), (\id_2, \del_2, \eps_2),..., (\id_n, \del_n, \eps_n) \}$ is available to all genuine ECUs on the network.
The delay is computed as the difference between two consecutive timestamps for the same ID in order to remove potential clock-skews. Indeed, by experimental measurements we did determine that clock skews may impede correct identification of delays. Concrete values for practical instances of the scheme and results are discussed next.

\subsection{Adversary model}

We consider the regular type of Dolev-Yao adversary that has full control over the communication channel. We note however that removing a genuine node from the bus is not easy in automotive-based scenarios since this requires either physical intervention or placing the node in the
bus-off state. The later possibility has been recently demonstrated
 in \cite{Cho16a} by exploiting the CAN error management system but it requires active error flags (which are visible on the bus) and the genuine node will remain in the bus-off state only for a fixed period of time. Moreover, removing a node will likely result in losing many functionalities from the car and many IDs from the bus, a behaviour which will likely be immediately recognized by the remaining ECUs. If the adversary cannot remove the genuine node from the bus it is very likely that the only effect of the adversary's intervention will be a DoS since injected frames will likely cause delays even for genuine frames. Addressing such a situation may be future work for us, but due to the design of the CAN bus a DoS will be always feasible, e.g., the adversary could simply corrupt any frame by overwriting one recessive bit with a dominant bit (this in turn may lead to a CRC error and all nodes will respond with active error flags, etc.). 

Assuming that the adversary can by some mean target a genuine node or a specific ID and remove it from the bus, the adversary further has to guess the exact delay at which the genuine frame needs to be sent. The success rate of an adversary can be estimated synthetically as:

\newcommand\aadv{\epsilon_{\mathit{adv}}}
\newcommand\aecu{\epsilon_{\mathit{ecu}}}

$$
\epsilon_{\mathit{adv}} = \frac{\tol}{2^\ell}
$$

Here $\tol$ is the delay tolerance for accepting a frame and $\ell$ is the security level (this are part of the protocol description that follows). This models the expected scenario where an adversary can at best insert a frame at some random point that hopefully will match the expected delay. 

\subsection{Results}

\begin{figure}[t!]
\scriptsize
\centering
\begin{minipage}{4.25cm}
\includegraphics[width=4.25 cm]{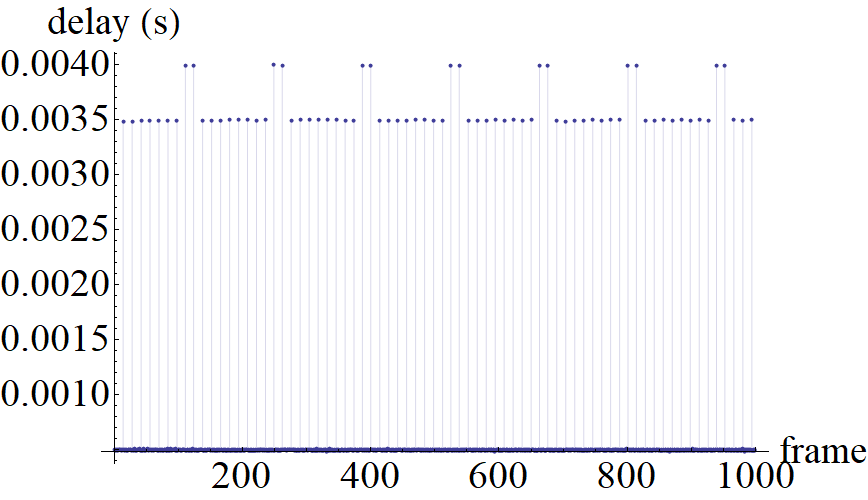}
\end{minipage}
\begin{minipage}{4.25cm}
\includegraphics[width=4.25 cm]{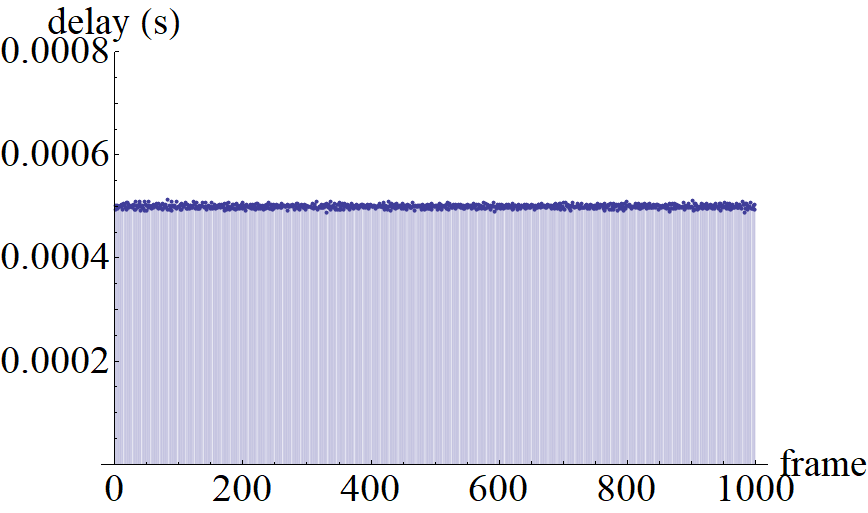}
\end{minipage}

(i)

\begin{minipage}{4.25cm}
\includegraphics[width=4.25 cm]{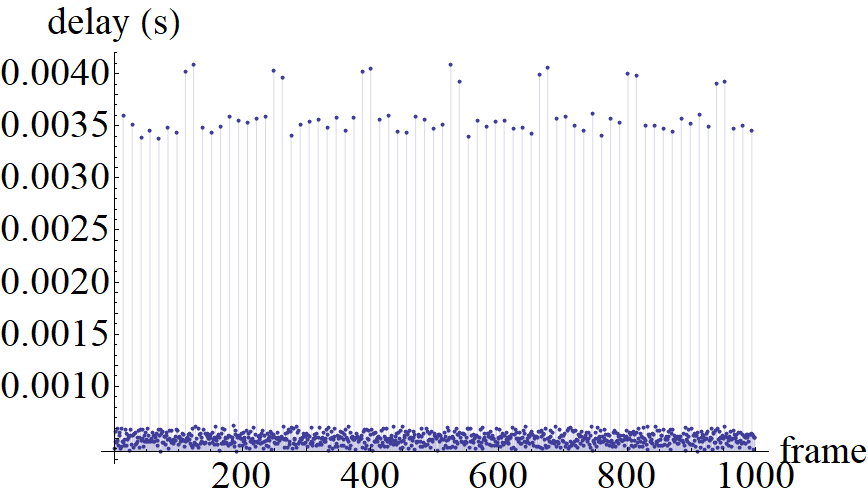}
\end{minipage}
\begin{minipage}{4.25cm}
\includegraphics[width=4.25 cm]{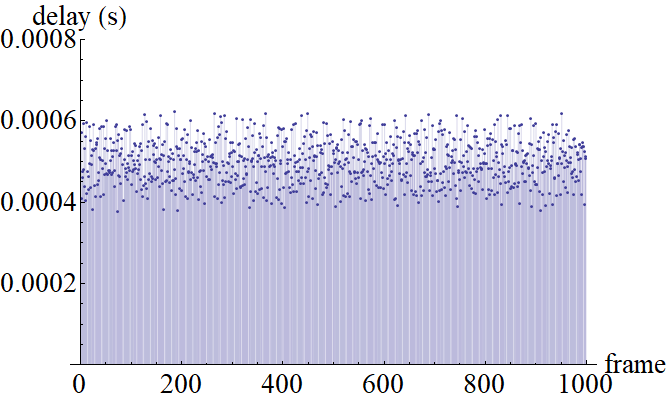}
\end{minipage}

(ii)

\caption{Interframe delays  (broadcast from Infineon TriCore node) in case of circular GCD optimization $\epsilon 0.5$ without a covert channel (i) and with the covert channel in place (ii)}
\label{fig:delays_gcd05}
\end{figure}

In Figure \ref{fig:delays_gcd05} we give an overview on the inter-frame delays on the bus with (ii) and without (i) the covert channel in place. Note that when the covert channel is in place, the delays vary with $\pm 128 \mu s$. This becomes more evident in the details from the right side of the picture for the case of frames separated by only $500 \mu s$.

The experiments that we carried proved to be consistent for all the IDs, regardless of the delays at which they are sent, i.e., $10, 20, 50$ or $100$ms. The variation of the estimated delay was in the order of $\pm 10 \mu s$ which is consistent with the time of 5 CAN bits (at 500kbps the duration of one bit on the CAN bus $\approx 2 \mu s$). This variation may be due to the variation in frame length due to the number of stuffing bits which differs (we improve on this next). 

\begin{figure}[thb!]
\scriptsize
\centering
\begin{minipage}{4.25cm}
\includegraphics[width=4.25 cm]{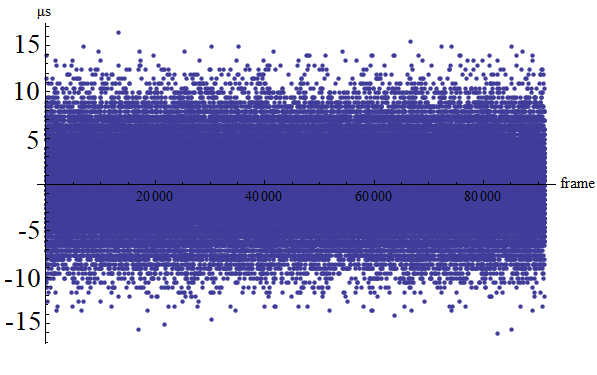}
\end{minipage}
\begin{minipage}{4.25cm}
\includegraphics[width=4.25 cm]{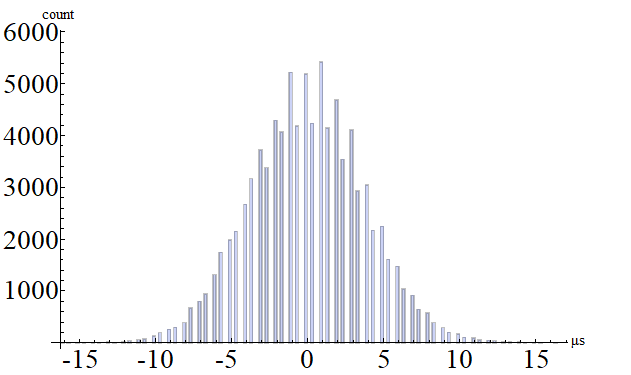}
\end{minipage}

(i)

\begin{minipage}{4.25cm}
\includegraphics[width=4.25 cm]{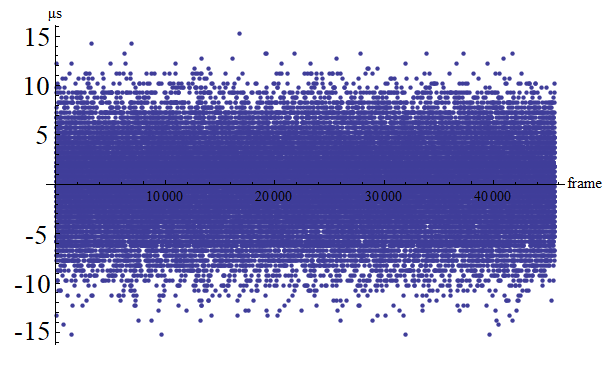}
\end{minipage}
\begin{minipage}{4.25cm}
\includegraphics[width=4.25 cm]{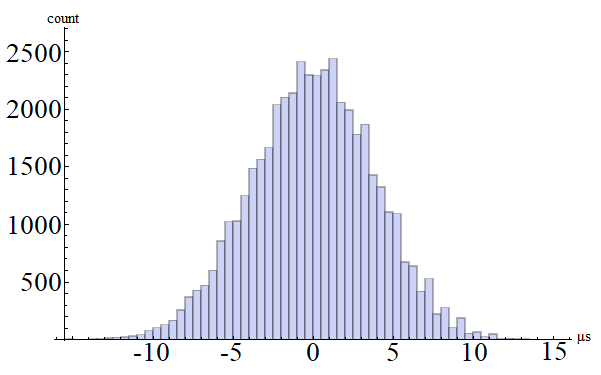}
\end{minipage}

(ii)

\begin{minipage}{4.25cm}
\includegraphics[width=4.25 cm]{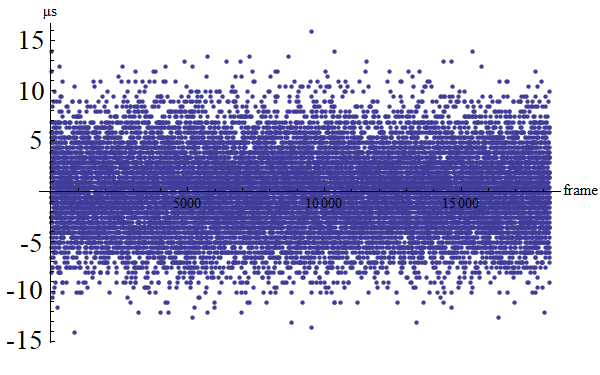}
\end{minipage}
\begin{minipage}{4.25cm}
\includegraphics[width=4.25 cm]{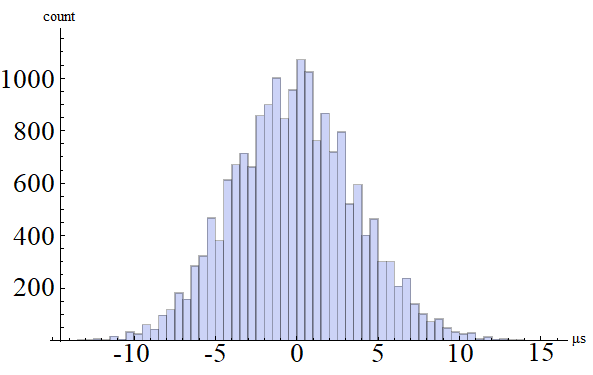}
\end{minipage}

(iii)

\begin{minipage}{4.25cm}
\includegraphics[width=4.25 cm]{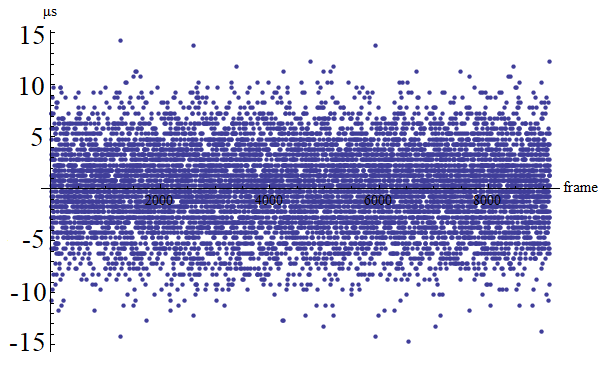}
\end{minipage}
\begin{minipage}{4.25cm}
\includegraphics[width=4.25 cm]{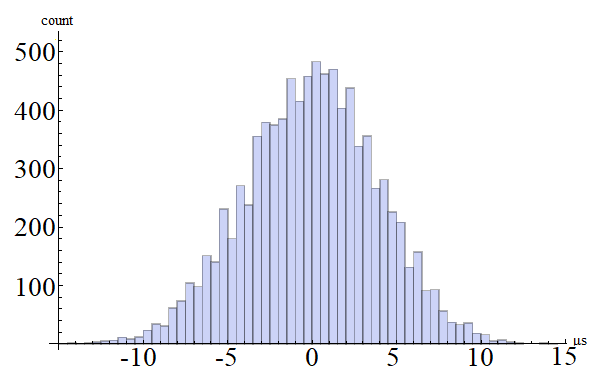}
\end{minipage}

(iv)

\caption{Experimental measurements for covert authenticated frames from an Infineon TriCore node: deviation from the expected delays (left) and histogram distribution (right) for an ID sent at $10 ms$ (i), $20 ms$ (ii), $50 ms$ (iii) and $100 ms$ (iv)}
\label{fig:delays_aut}
\end{figure}

Figure \ref{fig:delays_aut} shows the delays and their histogram distribution for frames broadcast at at $10 ms$ (i), $20 ms$ (ii), $50 ms$ (iii) and $100 ms$ (iv). Note that there are fewer samples as the delay increases. The deviation from the expected arrival time remains in the aforementioned range of  $\pm 10 \mu s$ for all frames and IDs. This is a  good result considering the busload which is identical to real-world operation of the CAN bus.  

\begin{figure}[thb!]
\scriptsize
\centering
\begin{minipage}{4.25cm}
\includegraphics[width=4.25 cm]{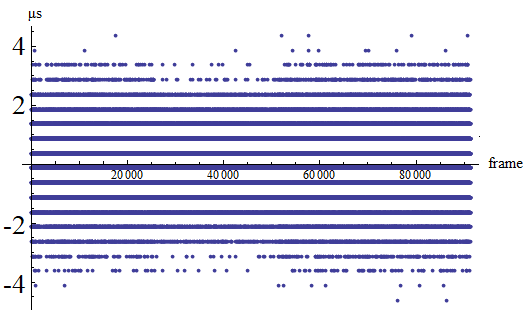}
\end{minipage}
\begin{minipage}{4.25cm}
\includegraphics[width=4.25 cm]{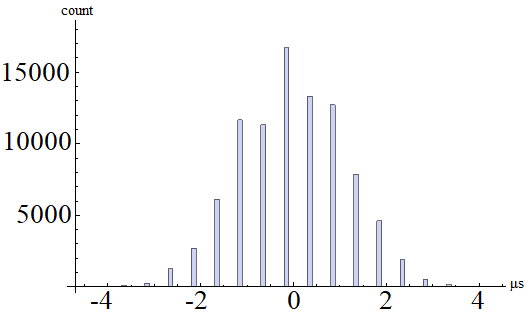}
\end{minipage}

(i)

\begin{minipage}{4.25cm}
\includegraphics[width=4.25 cm]{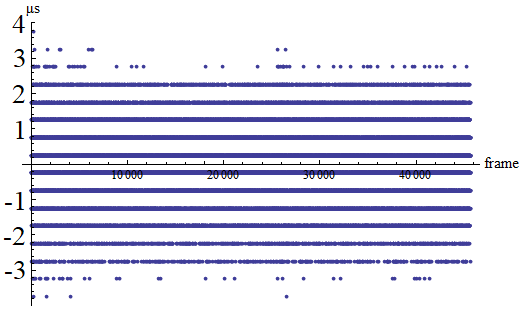}
\end{minipage}
\begin{minipage}{4.25cm}
\includegraphics[width=4.25 cm]{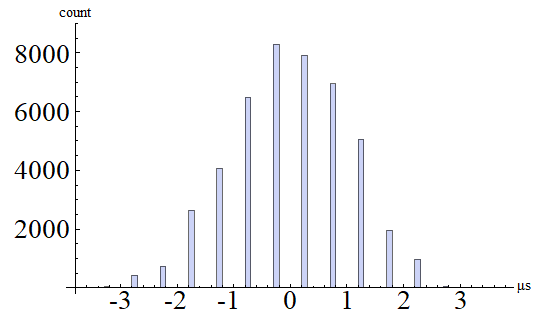}
\end{minipage}

(ii)

\begin{minipage}{4.25cm}
\includegraphics[width=4.25 cm]{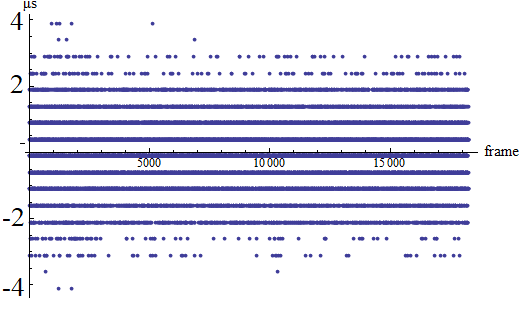}
\end{minipage}
\begin{minipage}{4.25cm}
\includegraphics[width=4.25 cm]{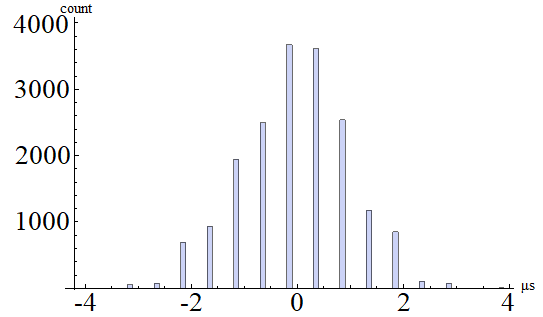}
\end{minipage}

(iii)

\begin{minipage}{4.25cm}
\includegraphics[width=4.25 cm]{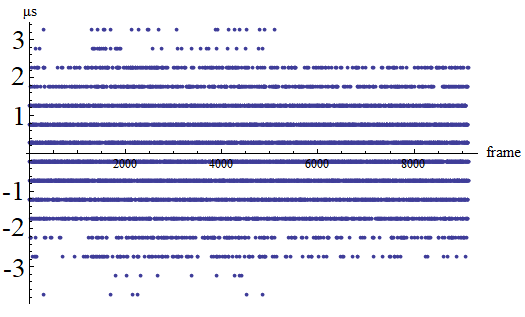}
\end{minipage}
\begin{minipage}{4.25cm}
\includegraphics[width=4.25 cm]{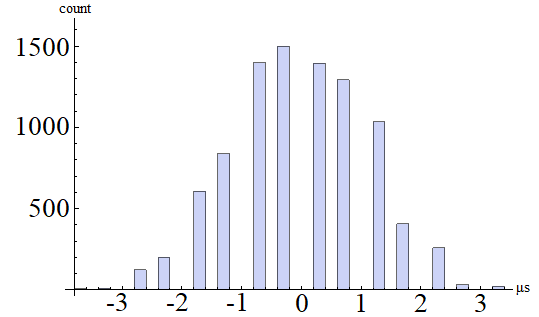}
\end{minipage}

(iv)

\caption{Experimental measurements for covert authenticated frames from an Infineon TriCore node: deviation from the expected delays (left) and histogram distribution (right) for an ID sent at $10 ms$ (i), $20 ms$ (ii), $50 ms$ (iii) and $100 ms$ (iv)}
\label{fig:delays_aut_2}
\end{figure}

To go even further, we have also taken the frame length into account and achieve a better match as depicted in Figure \ref{fig:delays_aut_2}.
The minimum error for all messages was at $-4.62 \mu s$ and the maximum at $4.87 \mu s$ which means that at a $5 \mu s$ tolerance all genuine messages will get the intended delay. 

In Table \ref{tab:res_tnr} we give the average rate for the true negatives given the $2, 3$ or $4 \mu s$ tolerance bound. We depict both the success rate for the genuine frames, i.e., $\aecu$, and the adversary advantage $\aadv$. In case of $\aecu$, the values are computed as the mean value taken over all the 40 IDs in a trace covering around 1.2 million frames. Concretely, the $2 \mu s$ error covered between $91.4 \% - 95.2 \% $ depending on the ID, while the $3 \mu s$ and $4  \mu s$ covered $99.21 \% - 99.93 \%$ and $99.96 \% - 100\%$ respectively. The adversary advantage $\aadv$ is synthetically computed as discussed previously. We also extend these results for the case of multiple frames, i.e., over $k$ frames, the adversary advantage becomes $\aadv^k$. The same happens for the advantage of the genuine frames which is now $\aecu^k$. In case when the $5 \mu s$ tolerance is used, all genuine frames are to be accepted, while the chance of an adversary to inject a frame is less than 1 in a million. Six frames are to be received in $1.5-6 ms$ as we discuss later, thus the authentication delay is not high.

\begin{table}[t!]
\centering
\caption{Success rates with tolerance $\epsilon \in \{2, 3, 4, 5 \} \mu s $ }
\label{tab:res_tnr}
\setlength{\tabcolsep}{0.5em}
\begin{tabular}{ l | c c c c c c c}

\hline
  $\tol$ & & 1 frame 	& 2 frames & 3 frames & 4 frames & 6 frames	  \\
   & \multicolumn{6}{c}{\%}  	  \\

\hline
\hline
\multirow{2}{*}{$2 \mu s$} 	&	$\aecu$ & 93.34  &  87.14 &  81.34    & 75.93  & 66.10 \\
												& $\aadv$ & 1.5  &  0.2 &  $0.0003$    & $< 10^{-5}$ & $<10^{-8}$  \\
\hline
\multirow{2}{*}{$3 \mu s$}	&	$\aecu$ 	& 99.56  &  99.12 & 98.68  & 98.25 & 97.38 \\
												&	$\aadv$ 	& 2.3  &  0.05 & 0.001  & $<10^{-4}$ & $< 10^{-7}$  \\
\hline
\multirow{2}{*}{$4 \mu s$}	&	$\aecu$  &  99.99 &  99.98 &  99.97   & 99.96 & 99.94 \\
												&	$\aadv$  &  3.1 &  0.09 &  0.002   & $<10^{-4}$ & $<10^{-7}$\\
\hline
\multirow{2}{*}{$5 \mu s$}	&	$\aecu$  &  100 &  100 &  100   & 100   & 100 \\
												&	$\aadv$	&  3.9 &  0.15 &  0.005   & $<10^{-3}$  &  $<10^{-6}$  \\

\hline												
\hline
\end{tabular}
\end{table}

\begin{figure}[t!]
\scriptsize
\centering
\begin{minipage}{4.25cm}
\includegraphics[width=4.25 cm]{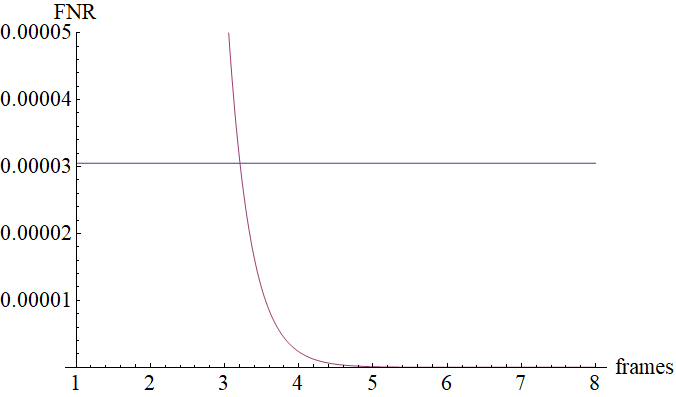}
\end{minipage}
\begin{minipage}{4.25cm}
\includegraphics[width=4.25 cm]{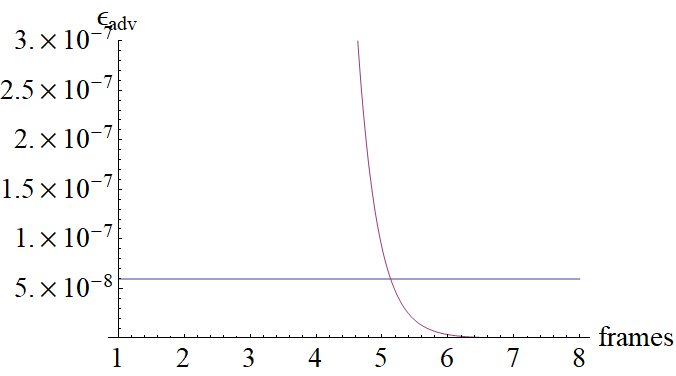}
\end{minipage}

\caption{Adversary success rate, i.e., $\aadv$, in 1--8 frames at $\rho = 5 \mu s$ ($\aecu=1.0$ all genuine frames accepted)}
\label{fig:adv_succ}
\end{figure}

In Figure~\ref{fig:adv_succ} we depict an estimation for the adversary success rate compared to an 15-bit security level (left)  and 24-bit security level (right). The 15-bit security level was chosen for comparison since it is the size of the CRC for CAN frames (the CRC is not resilient in front of adversaries but it worths as comparison), while the 24-bit security level is demanded by AUTOSAR \cite{AutosarSec}. The required security level of AUTOSAR can be reached in 5--6 frames. 

\begin{figure}[thb!]
\scriptsize
\centering
\begin{minipage}{4.25cm}
\includegraphics[width=4.25 cm]{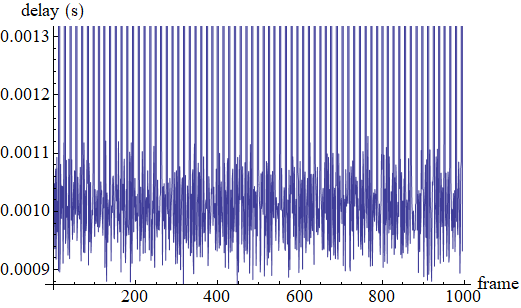}
\end{minipage}
\begin{minipage}{4.25cm}
\includegraphics[width=4.25 cm]{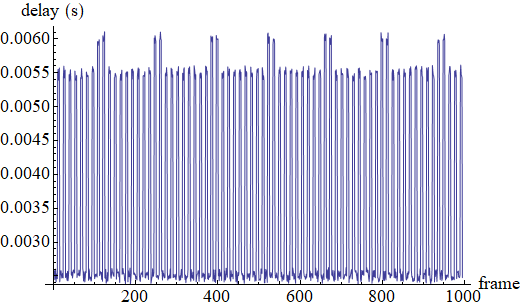}
\end{minipage}

\caption{Delay between 3 (left) or 6 (right) consecutive frames}
\label{fig:delays_36}
\end{figure}

\emph{Data-rate of the covert channel.} The data-rate of noisy channels can be computed with Arimoto-Blahut algorithms \cite{Arimoto72}, \cite{Blahut72}. To use these algorithms we have first extracted the channel matrix which gathers the probability that delay $\del' \in [1, 256] \mu s$ is encoded into delay $\del'' \in [1, 256] \mu s$. Once the channel was computed we used a freely available Matlab implementation of the algorithm\footnote{https://www.mathworks.com/matlabcentral/fileexchange/32757-channel-capacity-using-arimoto-blahut-algorithm} and determined that channel capacity is $\approx 4.9$ bits. 
Again, this suggests that the 24-bit security level can be achieved in six CAN frames.
This data-rate can be also estimated from the results in Table \ref{tab:res_tnr} since by setting a tolerance to $10 \mu s$ we get a noiseless channel. But now the $256$ symbols are reduced to $25$ and thus a data-rate of $\approx 4.6$ bits (this is a bit lower of the channel capacity computed with Arimoto-Blahut algorithms which gives an upper-bound).

\emph{Multi-frame authentication.} Cumulating delays over multiple frames is an option for increasing the security level. In 3 consecutive frames the security level is around 12 bits while for 6 consecutive frames it reaches the desired 24 bits for in-vehicle security. The authentication delay tops at $1.3 ms$ for 3 frames and $6 ms$ for 6 consecutive frames as depicted in Figure \ref{fig:delays_36}. This is just a worst case scenario since often the space between frames is $500 \mu s$ and thus around $1.5 ms$ or $3 ms$ are to be expected. This is a very small delay for the 24 bit authentication level considering that no cryptographic operation is needed except for the regular MAC and the bus-load is not increased.

\section{Conclusions}

We provide optimizations for scheduling CAN frames in order to reach an increased level of security on a time-covert cryptographic channel. The four optimization algorithms that we study show clear advantages for creating a covert timing channel on CAN. The Greedy Search and Circular GCD give better results, with a minimum inter-frame distance of 500 $\mu s$ which can further accommodate time-covert authentication. Over the optimized bus, the delays drift only in the $\pm 10 \mu s$ range which roughly corresponds to the the time of 5 CAN bits (the number of bits in each frame may drift due to stuffing bits). With these small deviations, around 4 bits can be covertly carried by each frame which means that the 24-bit security level demanded by recent standards may be reached in about 6 frames. Given the data rate and the fact that frames are spaced by around $250-500 \mu s$ this means that 24 security bits may be cumulated in around  $3ms$. Maintaining such strict delays is not necessarily easy, but we believe that modern microcontrollers can cope with them. Based on our analysis, the drifts from the expected arrival time are mostly caused by unoptimized traffic rather than by the accuracy of the controller's clock. Our procedures help in this direction by optimizing traffic allocation. Further investigations are needed to test the feasibility of the proposed procedures inside a real-world vehicle. Since modern time-triggered protocols, such as FlexRay, demand synchronization in the order of $10 \mu s$ in the worst case, we believe that covert channels may be implemented at the rigorous timing demands from our work. At the very least, we report an experimental upper bound, i.e., 4-5 bits, for such covert channels under optimized traffic flows.

~~\\

\textbf{Acknowledgement.} This work was supported by a grant of Ministry of Research
and Inovation, CNCS-UEFISCDI, project number PN-III-P1-1.1-TE-2016-1317, within
PNCDI III (2018-2020).

\bibliographystyle{abbrv}
\bibliography{cantis}

\newpage

\begin{IEEEbiography}[{\includegraphics[width=1in,height=1.25in,clip,keepaspectratio]{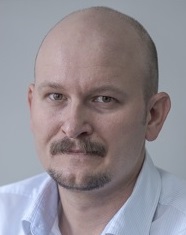}}]{Bogdan Groza} is Professor at Politehnica University of Timisoara (UPT). He received his Dipl.Ing. and Ph.D. degree from UPT in 2004 and 2008 respectively. In 2016 he successfully defended his habilitation thesis having as core subject the design of cryptographic security for automotive embedded devices and networks. He has been actively involved inside UPT with the development of laboratories by Continental Automotive and Vector Informatik. Besides regular participation in national and international research projects in information security, he lead the CSEAMAN project (2015-2017) and currently leads the PRESENCE project (2018-2019), two research programs dedicated to automotive security funded by the Romanian National Authority for Scientific Research and Innovation. 
\end{IEEEbiography}

\begin{IEEEbiography}[{\includegraphics[width=1in,height=1.25in,clip,keepaspectratio]{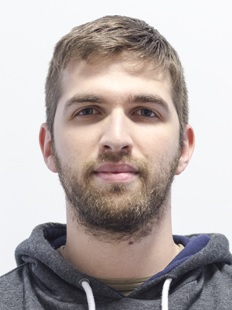}}]{Lucian Popa} started his PhD studies in 2018 at Politehnica University of Timisoara (UPT). He graduated his B.Sc in 2015 and his M.Sc studies in 2017 at the same university. He has a background of 4 years as a software developer and later system engineer in the automotive industry as former employee of Autoliv (2014 - 2018) and current employee of Veoneer (2018 - present). His research interests are in automotive security with focus on the security of in-vehicle buses.
\end{IEEEbiography}

\begin{IEEEbiography}[{\includegraphics[width=1in,height=1.25in,clip,keepaspectratio]{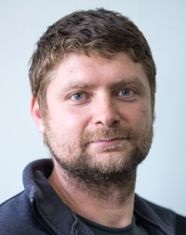}}]{Pal-Stefan Murvay} is Lecturer at Politehnica University of Timisoara (UPT). He graduated his B.Sc and M.Sc studies in 2008 and 2010 respectively and received his Ph.D. degree in 2014, all  from  UPT. He has a 10-year background as a software developer in the automotive industry. He worked as a postdoctoral researcher in the CSEAMAN project and is currently a senior researcher in the PRESENCE project. He also leads the SEVEN project related to automotive and industrial systems security. His current research interests are in the area of automotive security. 
\end{IEEEbiography} 

\end{document}